\documentclass[aps,prd,twocolumn,showpacs,superscriptaddress,nofootinbib,floatfix,showkeys,10pt]{revtex4-2}

\usepackage[colorlinks=true, linkcolor=blue, urlcolor=blue, citecolor=blue]{hyperref}
\usepackage{bm}
\usepackage{times}
\usepackage{braket}
\usepackage{amsfonts,amssymb,stmaryrd,latexsym,amsmath}
\usepackage[usenames,dvipsnames]{color}
\usepackage{slashed}
\usepackage{hyperref}
\usepackage{subfigure}
\usepackage{graphicx}
\usepackage{orcidlink}
\usepackage{multirow}
\usepackage{appendix}
\usepackage{dashrule}
\usepackage{float}
\allowdisplaybreaks[1]
\usepackage{verbatim}
\usepackage{overpic}

\newcommand{\sysu}{\affiliation{School of Physics, Sun Yat-sen University, Guangzhou 510275, China}}

\begin{document}
% \linenumbers
\title{Three-body molecular states composed of $D^{(*)}$ and two nucleons}

\author{Si-Yi Chen}\sysu

\author{Fei-Yu Chen}\sysu

\author{Xu-Liang Chen}\sysu

\author{Lu Meng \orcidlink{0000-0001-9791-7138}}\email{lmeng@seu.edu.cn}
\affiliation{School of Physics, Southeast University, Nanjing 211189, China }

\author{Ning Li\,\orcidlink{0000-0003-2987-2809}}\email{lining59@mail.sysu.edu.cn}\sysu
\author{Wei Chen\,\orcidlink{0000-0002-8044-5493}}\email{chenwei29@mail.sysu.edu.cn}
\sysu
\affiliation{Southern Center for Nuclear-Science Theory (SCNT), Institute of Modern Physics, 
Chinese Academy of Sciences, Huizhou 516000, Guangdong Province, China}

\begin{abstract}
We study the three-body systems $DNN$ and $D^{*}NN$ within a hadronic molecular framework by combining a realistic nucleon-nucleon interaction with a $D^{(*)}N$ potential constrained by heavy-quark symmetry. The three-body Schr\"odinger  equation is solved with the Gaussian Expansion Method, and the analytic structure of the spectrum is investigated using the Complex Scaling Method. We find that the $DNN$ system supports a robust and compact bound state in the $I(J^{P})=\tfrac{1}{2}(1^-)$ channel over a broad range of cutoff values, even when the corresponding $DN$ subsystem is weakly bound or unbound. For $D^{*}NN$,  the spin-$1$ nature of the heavy meson and the associated spin-dependent forces generate a clear spin hierarchy: deeply bound states appear in both $0^-$ and $2^-$ channels, while the $1^-$ channel exhibits a characteristic two-branch pattern with a strongly bound compact branch and a more weakly bound, spatially extended branch. The root-mean-square radii indicate pronounced spatial compression compared with the deuteron scale, highlighting the cooperative roles of realistic $NN$ correlations, the $D^{(*)}N$ interactions, and heavy-quark symmetry in forming compact heavy-flavor few-body bound states. No three-body resonances under complex scaling are found in the explored parameter space. Our results provide quantitative benchmarks for future experimental searches for such charmed-meson-nuclear bound states. 
\end{abstract}
\maketitle
\section{Introduction}\label{introduction}
In recent years, heavy flavor hadrons have provided a crucial testing ground for exploring the nonperturbative regime of quantum chromodynamics (QCD).  Experimentally, a number of exotic hadronic states containing charm or bottom quarks, have been observed, such as the $XYZ$ heavy-quarkonium-like states, $P_c$ pentaquark states and $T_{cc}$ tetraquark state. Representative experimental measurements can be found in Refs.~\cite{Belle:2003nnu,BESIII:2013ris,Belle:2013yex,LHCb:2015yax,LHCb:2019kea,LHCb:2021auc,LHCb:2021vvq}. For comprehensive reviews on this rapidly developing field, we refer the reader to Refs.~\cite{Swanson:2006st,Voloshin:2007dx,Brambilla:2010cs,Brambilla:2014jmp,Hosaka:2016pey,Chen:2016qju,Esposito:2016noz,Lebed:2016hpi,Ali:2017jda,Guo:2017jvc,Olsen:2017bmm,Liu:2019zoy,Brambilla:2019esw,Meng:2022ozq,Chen:2022asf,Liu:2024uxn,Wang:2025sic} . In this context, systems composed of open heavy-flavor mesons, such as $D$ and $D^*$, and nucleons have attracted considerable attention, as they provide a unique setting in which heavy-quark spin symmetry (HQSS) and light-quark chiral dynamics interplay, potentially supporting unconventional bound or resonant states.

At the two-body level, the nucleon--nucleon ($NN$) interaction has been studied extensively and is well constrained by experimental data~\cite{VANDERLEUN1982261,Horiuchi:2020bts,GREENWOOD1966702}. 
Precise measurements of scattering phase shifts, scattering lengths, and effective ranges at low energies have enabled the construction of high-precision phenomenological $NN$ potentials, such as the CD-Bonn, Argonne $v_{18}$, and Nijmegen models, as demonstrated in Refs.~\cite{Stoks:1994wp,Wiringa:1994wb,Park:1998cu,HASSANEEN2011566,Krebs:2020pii}. For broader overviews of the development and applications of modern $NN$ interactions, see the review articles in Refs.~\cite{Bethe:1940zz,Machleidt:2000ge,NPLQCD:2011naw}. These potentials typically incorporate meson-exchange mechanisms involving $\pi$, $\rho$, $\omega$, and $\sigma$ mesons, and successfully reproduce the properties of light nuclei, including the deuteron, thereby providing a solid microscopic foundation for nuclear force models. In the past few decades, significant progress has been made in constructing high-precision nuclear forces within chiral effective field theory~\cite{Epelbaum:2008ga,Machleidt:2011zz,Epelbaum:2019kcf,Krebs:2023gge,Krebs:2023ljo}.

In contrast to the well-established $NN$ interaction, the study of heavy-meson--nucleon systems, such as $DN$ and $D^{*}N$, is comparatively recent but has progressed rapidly in the past two decades. On the theoretical side, effective interactions constrained by heavy-quark symmetry and chiral symmetry for light quarks have been systematically developed. Early investigations based on chiral unitary and coupled-channel approaches demonstrated that near-threshold structures can be dynamically generated in the $DN$ and $D^{*}N$ systems~\cite{Lutz:2005ip,Garcia-Recio:2008rjt}. Subsequent studies employing one-boson-exchange models and effective Lagrangians further emphasized the role of tensor forces and channel coupling in providing attraction in selected quantum-number channels~\cite{He:2010zq,Haidenbauer:2010ch}. Recently, the $ND$ scattering lengths was extracted from the $\Lambda_b\to \pi^-pD^0$ decays by making use of the cusp at the $nD^+$ threshold.\cite{Sakai:2020psu}

HQSS implies a near degeneracy between the $D$ and $D^{*}$ mesons, which enhances the coupling between the $DN$ and $D^{*}N$ channels. As a result, the tensor component of the one-pion exchange potential plays a crucial role in generating attraction and enabling the possible formation of bound or resonant states, particularly in the isoscalar channels~\cite{Yasui:2009bz,Yamaguchi:2011qw,
Yamaguchi:2022oqz}. Within this framework, quark-model-based calculations, including the quark delocalization color screening model, have systematically studied the $DN$ and 
$D^{*}N$ systems and emphasized the role of channel coupling in producing binding~\cite{Zhao:2016zhf,Entem:2016lzh,Zhang:2019vqe}. Complementary studies based on effective field theory (EFT)
have provided a more model-independent description, predicting shallow bound or near-threshold states with sensitivity to short-range dynamics~\cite{Wang:2020dhf}. More recently, combined EFT and one-boson-exchange approaches with unitarization have further refined the description of the coupled $DN$--$D^{*}N$ dynamics and its connection to the 
charmed-baryon spectrum~\cite{Shen:2025akt}.

In this context, the $\Sigma_c(2800)$ and $\Lambda_c(2940)^+$ resonances have often been discussed as molecular candidates dominated by $DN$ and $D^{*}N$ components, respectively, 
motivated by their proximity to the corresponding thresholds%
~\cite{Dong:2010gu,He:2010zq,Ortega:2012cx,Zhang:2012jk}. Experimentally, direct constraints on heavy meson-nucleon interactions have begun to emerge only recently. The LHCb 
Collaboration has analyzed the amplitude structure of the $D^0 p$ system in decay $\Lambda_b^0 \to D^0 p \pi^-$, providing valuable information on the $DN$ interaction near threshold~\cite{LHCb:2017jym}, while the ALICE Collaboration has performed femtoscopic correlation measurements in proton--proton collisions, yielding the first quantitative probe of the $\bar{D}N$ interaction~\cite{ALICE:2022enj}.

Combining the well-established $NN$ potentials with the experimentally constrained $DN$ interactions naturally motivates the study of few-body systems containing a heavy meson and multiple nucleons. The three-body system $DNN$, with $D$ a charmed meson and $N$ a nucleon, provides the simplest nontrivial environment to explore how heavy-quark symmetry, realistic nuclear correlations, and meson-baryon dynamics cooperate in forming heavy-flavor few-body states~\cite{Garcilazo:2017ifi}.Beyond charm--nuclear systems, purely heavy-meson
three-body molecular configurations have also been
investigated, where heavy-quark symmetry and short-range
dynamics can generate nontrivial bound structures in
systems composed entirely of heavy mesons~\cite{Luo:2021ggs,Fu:2025joa,Wu:2025fzx}.
Earlier variational and Faddeev-type studies (including fixed-center approximations) pointed out that the $DNN$ system may support a quasibound structure with moderate binding and a narrow width in selected channels~\cite{Dote:2013zxa}. In the present work we go beyond this qualitative expectation by employing an energy-independent $D^{(*)}N$ interaction constrained by heavy-quark spin symmetry together with a realistic $NN$ potential, and by solving the three-body Schr\"odinger equation using the Gaussian Expansion Method (GEM)~\cite{Hiyama:2003cu}. Possible resonant structures are further analyzed within the Complex Scaling Method (CSM)~\cite{Aguilar:1971ve, Balslev:1971vb}. The combined application of GEM and CSM has proven highly effective in studying few-body resonant states across various scales, such as in few-hadron systems~\cite{Zhu:2024hgm,Wen:2025wit,Chen:2026ohg}, few-lepton systems~\cite{Ma:2025rvj,Wen:2025ehf}, and multiquark states~\cite{Chen:2023syh, Ma:2024vsi, Meng:2023jqk, Wu:2024euj, Wu:2024hrv, Wu:2024zbx}.

%We find that the $DNN$ system supports a deeply bound and compact state in the $I(J^{P})=\tfrac{1}{2}(1^-)$ channel over a wide range of model parameters, even when the corresponding two-body $DN$ subsystems are only weakly bound or unbound. For the $D^{*}NN$ system, the spin-1 nature of the $D^*$ meson enhances tensor and spin-dependent effects through coupled-channel dynamics, leading to deeply bound states and a clear spin hierarchy. The resulting rms distances~\cite{Hiyama:2012sma,Hiyama:2018ivm} indicate strong spatial compression compared with the deuteron scale, reflecting the cooperative roles of realistic $NN$ correlations, tensor forces, and heavy-quark spin symmetry in generating compact heavy-flavor few-body bound states.
It is also instructive to recall that few-body systems composed of a meson and two nucleons have been extensively investigated in the strange sector. In particular, the $\bar{K}NN$ ($K^-pp$) system has been studied using variational and Faddeev-type approaches, where the strong $\bar{K}N$ interaction can generate quasibound or deeply bound three-body states depending on the treatment of coupled channels and short-range dynamics~\cite{Faddeev:1960su,Akaishi:2002bg,Shevchenko:2007zz,Dote:2008in,Topolnicki:2015zya,Zhang:2021hcl}. Although the predicted binding energies and decay widths remain model-dependent, these studies illustrate a general mechanism: realistic $NN$ correlations combined with meson--nucleon attraction could produce nontrivial three-body structures and pronounced spatial rearrangement. Experimental indications of compact $\bar{K}NN$ configurations have also been reported in kaon-induced and proton-proton reactions~\cite{FINUDA:2005lqd,Yamazaki:2010mu}. This motivates analogous explorations in the charm sector, where the heavier meson mass reduces kinetic energy and heavy-quark spin symmetry enforces strong channel coupling between $DN$ and $D^{*}N$, potentially favoring the formation of bound states with distinctive spin dependence.

Motivated by these considerations, in this work we investigate the $DNN$ and $D^*NN$ systems in a unified potential-model framework. We construct the $D^{(*)}N$ interaction within the one-boson-exchange description consistent with HQSS and combine it with a realistic $NN$ interaction to solve the three-body bound-state problem using GEM, while CSM is employed to search for possible three-body resonances and clarify the analytic structure of the spectrum. We systematically analyze the dependence on model parameters, quantify the binding energies and spatial sizes, and elucidate the dynamical origin of the observed spin hierarchy (including the two-branch pattern in the $D^{*}NN$ $1^-$ channel). The results provide quantitative inputs for future experimental searches for heavy-flavor few-body states at facilities such as LHC, J-PARC, and GSI-FAIR, and help clarify the interplay between heavy-meson dynamics and nuclear correlations in few-body systems.

The article is organized as follows. After the introduction, in section~\ref{formalism} we present the formalism including the 
$NN$ interaction, $DN$ interaction, and the Gaussian Expansion Method as well as the Complex Scaling Method. We give the numerical results and perform detailed analysis in section~\ref{results}. In the last section~\ref{conclusion}, we
summarize the results and draw a conclusion. 

\section{Formalism}\label{formalism}
In this section, we present the theoretical framework employed to investigate the three-body $D^{(*)}NN$ systems. We first introduce the Hamiltonian and describe the interaction potentials for the $NN$ and $D^{(*)}N$ subsystems. We then outline the GEM adopted to solve the three-body Schrödinger equation and the CSM used to analyze bound and resonant states.

The three-body system $DNN$ is described by the nonrelativistic Hamiltonian
\begin{equation}
H = T + V_{NN} + V_{D^{(*)}N}^{(1)} + V_{D^{(*)}N}^{(2)}\, ,
\end{equation}
where $T$ denotes the kinetic energy operator with the center-of-mass motion removed, $V_{NN}$ represents the nucleon-nucleon interaction, and $V_{D^{(*)}N}^{(i)}$ ($i=1,2$) denotes the interaction between the $D^{(*)}$ meson and the $i$-th nucleon. Three-body forces are neglected in the present study.

\begin{figure}
  \centering
  \includegraphics[width=0.5\textwidth]{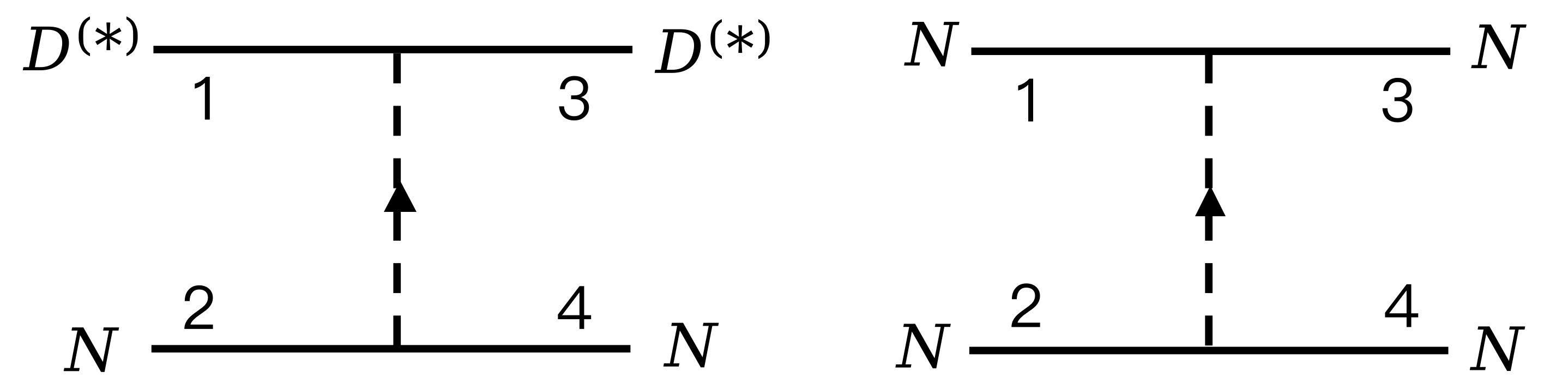}
  \caption{Feynman diagrams used to derive OBE potentials for the $D^{(*)}N$ and $NN$ systems.}
  \label{fig:feynman}
\end{figure}
%%%%%%%%%%%%%%%%%%%%%%%%%%%%%%%%%%%%%%%%%
\subsection{Nucleon-nucleon interaction}
For the nucleon-nucleon interaction, we employ the CD-Bonn potential, which is based on a relativistic one-boson-exchange framework and provides a high-precision description of $NN$ scattering data and deuteron properties~\cite{Machleidt:1987hj}. In this model, the $NN$ interaction is generated from an effective meson-nucleon Lagrangian involving the exchange of pseudoscalar, scalar, and vector mesons, namely $\pi$, $\sigma$, $\eta$, $\rho$, and $\omega$.

The interaction Lagrangian density can be written schematically as
\begin{equation}
\mathcal{L}_{NN} = \mathcal{L}_{\pi NN} + \mathcal{L}_{\sigma NN} + \mathcal{L}_{\rho NN} + \mathcal{L}_{\omega NN}+ \mathcal{L}_{\eta NN},
\end{equation}
with the individual interaction terms given by
\begin{align}
\mathcal{L}_{\pi NN} &= g_{\pi NN}\bar{N} i\gamma_5 \vec{\tau}  \cdot \vec{\pi} N,\notag \\
\mathcal{L}_{\sigma NN} &= g_{\sigma NN} \bar{N} \sigma N,\notag \\
\mathcal{L}_{\rho NN} &= g_{\rho NN} \bar{N} \gamma^\mu \vec{\tau}  \cdot \vec{\rho}_\mu N \notag \\
 & +\frac{f_{\rho NN}}{4M_N} \bar{N} \sigma^{\mu\nu} \vec{\tau}  \cdot \left(\partial_\nu \vec{\rho}_\mu - \partial_\mu \vec{\rho}_\nu\right)N,\notag \\
\mathcal{L}_{\omega NN} &= g_{\omega NN} \bar{N} \gamma^\mu  \omega_\mu N,
\end{align}
where $N$ denotes the nucleon field, $\vec{\tau}$ are the Pauli matrices for isospin  and $M_N$ is the nucleon mass. The coupling constants are chosen consistently with the CD-Bonn parametrization. In the present calculation, the coupling constants and isospin operators for the $NN$ interaction are taken from Table~\ref{tab:coupling_Oiso}. Due to the Pauli exclusion principle for identical fermion system, only the deuteron channel with total isospin $I=0$ is considered, and the nucleon mass is fixed as $m_N = 939~\mathrm{MeV}$.

\begin{table*}[htbp]
\centering
\renewcommand{\arraystretch}{2.4}
\setlength{\tabcolsep}{8pt}
\caption{Parameters of the one-boson-exchange interactions for the $D^{(*)}N$ and $NN$ systems, including 
the masses of the exchanged bosons~\cite{ParticleDataGroup:2022pth}, the corresponding coupling 
constants~\cite{Zhu:2024hgm,Machleidt:2000ge}, and the isospin operator $O_{I}$. $\vec{\tau}_i(i = 1, 2)$ 
is the Pauli matrix for the isospin of nucleon $i$, and satisfies $\vec{\tau}_1 \cdot \vec{\tau}_2 = -3$ for 
total isospin $I=0$ and $\vec{\tau}_1 \cdot \vec{\tau}_2 = 1$ for $I=1$. 
The axial coupling is taken as $g = 0.59$ while the pion decay coupling is $f_\pi = 0.132$ GeV. $\lambda$ 
is given in units of GeV$^{-1}$. We note that the $\sigma$ coupling is important but remains the most uncertain quantity at present. 
Its strength is expected to be derived from theoretical frameworks and ultimately from first-principles QCD in the future. In this work, 
we take its value as a baseline and redetermine it using the experimental pole position of the $Z_c$ state. 
In contrast to the well-established coupling constants for the $NN$ interaction, 
those for the $D^{(*)}N$ systems are model-dependent. The values adopted here serve merely as a baseline, 
with variations around this baseline accounted for in Eq.~\eqref{eq:couplingrescale} and the surrounding discussion. }
\label{tab:coupling_Oiso}
\begin{tabular}{c c c c c c c c c c}
\hline\hline
\multirow{2}{*}{Meson}
& \multirow{2}{*}{Mass (MeV)}
& \multicolumn{6}{c}{$C_{coupling}$}
& \multicolumn{2}{c}{$O_{I}$} \\
\cline{3-10}
& 
& $g_s$
& $\beta g_V$
& $\lambda g_V$
& $\dfrac{g}{f_\pi}$
& $\dfrac{g_{\alpha NN}^2}{4\pi}$
& $\dfrac{f_{\alpha NN}}{g_{\alpha NN}}$
& \parbox[c]{2.6cm}{\centering $NN$}
& \parbox[c]{2.6cm}{\centering $DN$} \\
\hline
$\sigma$ & 500 & 0.76 & -- & -- & -- & 7.78 & -- 
& \parbox[c]{2.6cm}{\centering $I$}
& \parbox[c]{2.6cm}{\centering $I$} \\
$\pi$ & 137 & -- & -- & -- & 4.47 & 14.9 & -- 
& \parbox[c]{2.6cm}{\centering $\dfrac{\vec{{\tau}_1}\!\cdot\!\vec{{\tau}_2}}{2}$}
& \parbox[c]{2.6cm}{\centering $-\dfrac{\vec{{\tau}_1}\!\cdot\!\vec{{\tau}_2}}{2}$} \\
$\eta$ & 548 & -- & -- & -- & 4.47 & 3 & -- 
& \parbox[c]{2.6cm}{\centering $\dfrac{1}{6} I$}
& \parbox[c]{2.6cm}{\centering $\dfrac{1}{6} I$} \\
$\rho$ & 775 & -- & 5.22 & $3.25$ & -- & $0.95$ & $6.1$
& \parbox[c]{2.6cm}{\centering $\dfrac{\vec{{\tau}_1}\!\cdot\!\vec{{\tau}_2}}{2}$}
& \parbox[c]{2.6cm}{\centering $-\dfrac{\vec{{\tau}_1}\!\cdot\!\vec{{\tau}_2}}{2}$} \\
$\omega$ & 783 & -- & 5.22 & $3.25$ & -- & 20 & 0
& \parbox[c]{2.6cm}{\centering $\dfrac{1}{2}I$}
& \parbox[c]{2.6cm}{\centering $\dfrac{1}{2}I$} \\
\hline\hline
\end{tabular}
\end{table*}

After performing a nonrelativistic reduction and Fourier transformation to coordinate space, the resulting $NN$ interaction can be expressed in terms of a set of effective potential operators. In particular, four independent operators are introduced to describe the central, pseudoscalar, tensor, and spin--orbit components of the interaction:
\begin{equation}
\begin{aligned}
\hat{\mathcal{V}}_{\mathrm{C}} & \equiv  \delta _{1,3} \delta _{2,4},\\
\hat{\mathcal{V}}_{\mathrm{P}} & \equiv \tfrac{1}{3} (\vec{\sigma }_{1,3} \cdot \vec{\sigma }_{2,4} )\hat{\mathcal{P}} 
+\tfrac{1}{3}\hat{T} (\vec{\sigma }_{1,3} ,\vec{\sigma }_{2,4} )\hat{\mathcal{Q}},\\
\hat{\mathcal{V}}_{\mathrm{T}} & \equiv \tfrac{2}{3} (\vec{\sigma }_{1,3} \cdot \vec{\sigma }_{2,4} )\hat{\mathcal{P}} 
-\tfrac{1}{3}\hat{T} (\vec{\sigma }_{1,3} ,\vec{\sigma }_{2,4} )\hat{\mathcal{Q}},\\
\hat{\mathcal{V}}_{\mathrm{LS}} & \equiv 
\left[\vec{L} \cdot \vec{\sigma }_{1,3} \delta _{2,4} 
+\delta _{1,3} \vec{L} \cdot \vec{\sigma }_{2,4}\right ]\hat{\mathcal{R}}.
\end{aligned}
\end{equation}
Here the indices $(1,2)$ and $(3,4)$ refer to the nucleons in the initial and final states , respectively as illustrated in Fig.~\ref{fig:feynman}. The operators $\vec{\sigma}_{1,3}$ and $\vec{\sigma}_{2,4}$ represent the Pauli spin
matrices acting on the corresponding initial--final nucleon spin spaces, and the Kronecker deltas $\delta_{1,3}$ and $\delta_{2,4}$ enforce the matching of particle labels between the initial and final states.The orbital angular momentum operator $\vec{L}$ is defined with respect to the relative coordinate of the interacting hadrons. The radial differential operators are defined as
\begin{equation}
\begin{aligned}
\hat{\mathcal{P}} \equiv r^{-2} \partial _{r} (r^{2} \partial _{r} ),\quad 
\hat{\mathcal{Q}} \equiv r\partial _{r} (r^{-1} \partial _{r} ),\quad 
\hat{\mathcal{R}} \equiv r^{-1} \partial _{r}\, ,
\end{aligned}
\end{equation}
and the tensor operator is given by
\begin{equation}
\begin{aligned}
\hat{T}(\vec{A} ,\vec{B}) & \equiv 3(\vec{A} \cdot \hat{r})(\vec{B} \cdot \hat{r}) -\vec{A} \cdot \vec{B}\, .
\end{aligned}
\end{equation}
Using the above operators, the meson-exchange contributions to the $NN$ interaction can be written in a compact form as
\begin{equation}
\begin{aligned}
V^{\mathrm{\sigma}} &= -\tfrac{g_{\sigma NN}^{2}}{4\pi}
\left[ \left(1+\tfrac{1}{4m_N^{2}}\hat{\mathcal{P}} \right)\hat{\mathcal{V}}_{\mathrm{C}} 
-\tfrac{1}{4m_N^{2}}\hat{\mathcal{V}}_{\mathrm{LS}}\right]\mathcal{V}_{\mathrm{\sigma}},\\
V^{\mathrm{v}} &=
\begin{aligned}[t]
&\tfrac{g_{v NN}^{2}}{4\pi}
\left[\hat{\mathcal{V}}_{\mathrm{C}}
-\tfrac{1}{4m_N^{2}}\hat{\mathcal{V}}_{\mathrm{LS}}
-\tfrac{1}{4m_N^{2}}\hat{\mathcal{V}}_{\mathrm{T}}\right]H_{v} \\
&-4\left(\tfrac{f_{v NN}}{4m_N}\right)^{2} \hat{\mathcal{V}}_{\mathrm{T}} H_{v}
- 2\tfrac{f_{v NN} g_{v NN}}{m_N}
\left[ \hat{\mathcal{P}} \hat{\mathcal{V}}_{\mathrm{C}} - \hat{\mathcal{V}}_{\mathrm{T}}  \right]H_{v},
\end{aligned}\\
V^{\mathrm{p}} &= -\tfrac{g_{\pi NN}^{2}}{16m_N^{2}}\hat{\mathcal{V}}_{\mathrm{P}}H_{p}.
\end{aligned}
\end{equation}
The functions $H_{p}$ and $H_{v}$ are defined as
\begin{align}
H_{v}&= O_I \mathcal{V}_{\rho} +O_I \mathcal{V}_{\omega },\\
H_{p}&=O_I\mathcal{V}_{\pi } + O_I\mathcal{V}_{\eta },
\end{align}
where the isospin-related operators $O_I$ are given in Table~\ref{tab:coupling_Oiso}, and the radial functions $\mathcal{V}_{\alpha}$ take the form
\begin{equation}
\mathcal{V}_{\alpha} = \tfrac{e^{-m_\alpha r}} {4\pi r} - \tfrac{e^{-\Lambda r}} {4\pi r} - \tfrac{\Lambda^2-m_\alpha^2}{8\pi \Lambda}e^{-\Lambda r}.
\end{equation}

To account for the finite size of hadrons, a monopole form factor
\begin{equation}
F_\alpha(q) = \frac{\tilde{\Lambda}^2 - m_\alpha^2}{\tilde{\Lambda}^2 - q^2},~\label{eq:formfactor}
\end{equation}
is introduced at each interaction vertex in the $NN$ interaction. Here $\tilde{\Lambda}$ is used for the $NN$ interaction. 
The same monopole form factor is adopted for the $D^{(*)}N$ interaction, with an independent cutoff parameter $\Lambda$.

%%%%%%%%%%%%%%%%%%%%%%%%%%%%%
\subsection{\texorpdfstring{$DN$}{DN} interaction}
With the heavy quark spin symmetry, the pseudoscalar meson $D$ and vector meson $D^*$ can form a superfield $\mathcal{H}$ ~\cite{Casalbuoni:1996pg,Falk:1992cx,Wise:1992hn,Li:2012ss,Liu:2008xz}:

\begin{equation}
\mathcal{H} = \frac{1 + \not{v}}{2} \bigl( P_\mu^* \gamma^\mu - P \gamma_5 \bigr),
\end{equation}
where $P = (D^0, D^+)$ and $P_\mu^* = (D^{*,0}, D^{*,+})_\mu$, and $v^\mu = (1, \mathbf{0})$ is the four-velocity of the heavy meson. 

%Similarly, the antiparticles can be represented by the superfield $\tilde{\mathcal{H}}$:
%\begin{equation}
%is \tilde{\mathcal{H}} = \bigl( \tilde{P}_\mu^* \gamma^\mu - \tilde{P} \gamma_5 \bigr) \frac{1 - \not{v}}{2},
%\end{equation}
%with $\tilde{P} = (\bar{D}^0, D^-)^{\mathrm{T}}$ and $\bar{P}_\mu^* = (\bar{D}^{*0}, D^{*-})_\mu^{\mathrm{T}}$. We adopt the charge conjugation convention: $D \xrightarrow{C} \bar{D}$, $D^* \xrightarrow{C} -\bar{D}^*$, i.e., $\mathcal{H} \xrightarrow{C} C^{-1} \tilde{\mathcal{H}}^{\mathrm{T}} C$, where $C = i\gamma^2\gamma^0$ is the charge conjugation matrix. 
The Hermitian conjugate of $\mathcal{H}$
is defined as 
%and $\tilde{\mathcal{H}}$ are defined as 
$\bar{\mathcal{H}} = \gamma_0 \mathcal{H}^\dagger \gamma_0$.
%and $\bar{\tilde{\mathrm{H}}} = \gamma_0 \tilde{\mathcal{H}}^\dagger \gamma_0$, respectively.
In the one-boson-exchange model, the Lagrangian reads as 
\begin{equation}
\begin{aligned}
\mathcal{L} =\;&
g_s \, \mathrm{Tr}\!\left[ \mathcal{H} \sigma \bar{\mathcal{H}} \right]
+ i g \, \mathrm{Tr}\!\left[ \mathcal{H} \gamma_\mu \gamma_5 A^\mu \bar{\mathcal{H}} \right] \\
&+ i \beta \, \mathrm{Tr}\!\left[ \mathcal{H} v_\mu (\mathcal{V}^\mu - \rho^\mu) \bar{\mathcal{H}} \right]
+ i \lambda \, \mathrm{Tr}\!\left[ \mathcal{H} \sigma_{\mu\nu} F^{\mu\nu} \bar{\mathcal{H}} \right], \\
%&+ g_s \, \mathrm{Tr}\!\left[ \bar{\tilde{\mathcal{H}}} \sigma \tilde{\mathcal{H}} \right]
%+ i g_a \, \mathrm{Tr}\!\left[ \bar{\tilde{\mathcal{H}}} \gamma_\mu \gamma_5  \mathcal{A}^\mu \tilde{\mathcal{H}} \right] \\
%&- i \beta \, \mathrm{Tr}\!\left[ \bar{\tilde{\mathcal{H}}} v_\mu (\mathcal{V}^\mu - \rho^\mu) \tilde{\mathcal{H}} \right]
%+ i \lambda \, \mathrm{Tr}\!\left[\bar{\tilde{\mathcal{H}}}   \sigma_{\mu\nu} F^{\mu\nu} \tilde{\mathcal{H}} \right] .
\end{aligned}
\end{equation}
where $F^{\mu\nu} = \partial^\mu \rho^\nu - \partial^\nu \rho^\mu - [ \rho^\mu , \rho^\nu ]$ is the field strength tensor of the vector mesons, and $\mathcal{V}^\mu$ and $\mathcal{A}^\mu$ are the vector and axial-vector building blocks of the pseudoscalar mesons, respectively:
\begin{equation}
\mathcal{V}^\mu = \frac{1}{2} \bigl[ \xi^\dagger\partial^\mu \xi \bigr],
\mathcal{A}^\mu = \frac{1}{2} \bigl\{ \xi^\dagger \partial^\mu \xi \bigr\},
\end{equation}
with 
\begin{equation}
\xi = \exp\!\bigl( i \mathbb{P} / f_\pi \bigr).
\end{equation}
The pseudoscalar meson matrix $\mathbb{P}$ is given by:
\begin{equation}
\mathbb{P} = 
\begin{pmatrix}
\displaystyle \frac{\pi^0}{\sqrt{2}} + \frac{\eta}{\sqrt{6}} & \pi^+ \\[6pt]
\pi^- & \displaystyle -\frac{\pi^0}{\sqrt{2}} + \frac{\eta}{\sqrt{6}}
\end{pmatrix},
\end{equation}
and the vector meson multiplet $\rho^\mu$ is defined as:
\begin{equation}
\rho^\mu = \frac{i g_V}{\sqrt{2}}
\begin{pmatrix}
\displaystyle \frac{\rho^0 + \omega}{\sqrt{2}} & \rho^+ \\[6pt]
\rho^- & \displaystyle -\frac{\rho^0 + \omega}{\sqrt{2}}
\end{pmatrix}^{\!\!\mu}.
\end{equation}

Starting from the effective Lagrangian given above, the $DN$ interaction employed in this work is constructed within an effective meson-exchange framework constrained by heavy-quark spin symmetry and chiral symmetry. After a nonrelativistic reduction, we obtain an energy-independent interaction potential in the coordinate space,  whose operator structure and radial dependence are consistent with those commonly used in the literature. The masses of mesons, coupling constants, and isospin operators entering the $D^{(*)}N$ interaction are summarized in Table~\ref{tab:coupling_Oiso}. The masses of the $D$ and $D^*$ mesons are taken as $m_D = 1867~\mathrm{MeV}$ and $m_{D^*} = 2009~\mathrm{MeV}$, respectively.

%Within this framework, the $DN$ and $D^{*}N$ interactions are treated as a coupled-channel system.
We now present the explicit forms of the effective potentials for the transitions $DN\to DN$ and $D^{*}N\to D^{*}N$ respectively, 
\begin{align}
\mathcal{V}_{DN\to DN} 
&= \Big[ g_s g_{\sigma NN}
\big( \hat{\mathcal{V}}_{1}^{\mathrm{C}}
- \tfrac{1}{4m_N^{2}} \hat{\mathcal{V}}_{1}^{\mathrm{LS}} \big)
\Big]  \mathcal{V}_{\sigma} \notag \\
&\quad - \Big[
\tfrac{\sqrt{2}}{2} \beta g_V g_{v NN}
\big( \hat{\mathcal{V}}_{1}^{\mathrm{C}}
+ \tfrac{1}{4m_N^{2}} \hat{\mathcal{V}}_{1}^{\mathrm{LS}} \big)\notag\\
&\quad + \tfrac{\sqrt{2}\,\beta g_Vf_{vNN}}{2m_N}
\big( \hat{\mathcal{V}}_{1}^{\mathrm{C}} \hat{\mathcal{P}}
+ 2\hat{\mathcal{V}}_{1}^{\mathrm{LS}} \big)
\Big] H_{v},
\end{align}
and
\begin{align}
\mathcal{V}_{D^*N\to D^*N} 
&= \Big[
g_s g_{\sigma NN}
\big( \hat{\mathcal{V}}_{2}^{\mathrm{C}}
- \tfrac{1}{4m_N^{2}} \hat{\mathcal{V}}_{2}^{\mathrm{LS}} \big)
\Big] \mathcal{V}_{\mathrm{\sigma}} \notag \\
&\quad + \Big[
\tfrac{2 g g_{\pi NN}}{4 f_\pi m_N}
\hat{\mathcal{V}}_{2}^{\mathrm{P}}
\Big] H_{p} \notag \\
&\quad - \Big[
\tfrac{\sqrt{2}}{2} \beta g_V g_{v NN}
\big( \hat{\mathcal{V}}_{2}^{\mathrm{C}}
+ \tfrac{1}{4m_N^{2}} \hat{\mathcal{V}}_{2}^{\mathrm{LS}} \big)\notag\\
&\quad+ \tfrac{\sqrt{2} \beta g_V f_{vNN}}{2m_N}
\big( \hat{\mathcal{V}}_{2}^{\mathrm{C}} \hat{\mathcal{P}}
+ 2\hat{\mathcal{V}}_{2}^{\mathrm{LS}} \big)
\Big] H_{v}\notag \notag\\
&\quad + \Big[
\tfrac{2\sqrt{2}\lambda g_V f_{v NN}}{4m_N}\hat{\mathcal{V}}_{2}^{\mathrm{T}}\notag\\
&\quad + \tfrac{2\sqrt{2}\lambda g_V g_{v NN}}{4m_N}
\big( \hat{\mathcal{V}}_{2}^{\mathrm{T}}
- 2\hat{\mathcal{V}}_{2}^{\mathrm{LS}} \big)
\Big] H_{v}.
\end{align}
We adopt the following operators, 
\begin{align}
&\hat{\mathcal V}^{C}_{1}\equiv \chi_1^\dagger \chi_2 ,  \quad\quad \quad \quad \quad \hat{\mathcal V}^{LS}_{1}\equiv \vec{L}\cdot \vec{\sigma}_{2,4} \hat{\mathcal R},\notag \\
&\hat{\mathcal V}^{C}_{2}\equiv 
\vec{\epsilon}_3^{\,*}\!\cdot \vec{\epsilon}_1\,
\chi_1^\dagger \chi_2 , \quad\quad\quad
\hat{\mathcal V}^{LS}_{2}\equiv 
\vec{\epsilon}_3^{\,*}\!\cdot \vec{\epsilon}_1
\vec{L}\cdot \vec{\sigma}_{2,4}\hat{\mathcal R},\\
&\hat{\mathcal V}^{P}_{2} \equiv \frac{1}{3}\left(
i\,\vec{\epsilon}_3^{\,*}\!\times \vec{\epsilon}_1 \cdot \vec{\sigma}_{2,4}
\right)\hat{\mathcal P}
\;+\;
\frac{1}{3}\,
\hat{T}\!\left(
i\,\vec{\epsilon}_3^{\,*}\!\times \vec{\epsilon}_1,\,
\vec{\sigma}_{2,4}
\right)\,\hat{\mathcal Q},\notag
\\
&\hat{\mathcal V}^{T}_{2}\equiv
\frac{2}{3}\left(
i\,\vec{\epsilon}_3^{\,*}\!\times \vec{\epsilon}_1 \cdot \vec{\sigma}_{2,4}
\right)\,\hat{\mathcal P}
\;-\;
\frac{1}{3}\,
\hat{T}\!\left(
i\,\vec{\epsilon}_3^{\,*}\!\times \vec{\epsilon}_1,\,
\vec{\sigma}_{2,4}
\right)\,\hat{\mathcal Q}.\notag
\end{align}
% Here, the vectors $\vec{\epsilon}_i$ $(i=1,2)$ and
% $\vec{\epsilon}_i^{\,*}$ $(i=3,4)$ denote the polarization vectors of the
% initial and final $D^*$ mesons, respectively.

Here, ${\vec \epsilon}_1({\vec \epsilon}_3^*)$ denotes the polarization vector of the incoming (outgoing) 
$D^*$ meson while ${\vec \sigma}_2({\vec \sigma}_4)$ denotes the 
spin Pauli matrix of the incoming (outgoing) nucleon.

The same monopole form factor as in Eq.~\eqref{eq:formfactor} is also employed for the $D^{(*)}N$ interaction, with an independently assigned cutoff parameter $\Lambda$. To consistently account for this regulator dependence, the vertex coupling constants are allowed to vary with $\Lambda$ ($\Lambda_{D^{(*)}N} \equiv \Lambda$). Following Refs.~\cite{zhu_2024_14464767,Zhu:2024hgm}, the DD vertex coupling constants are parametrized as
\begin{equation}
\lambda \rightarrow \lambda\, R_\lambda, \qquad
\beta \rightarrow \beta\, R_\beta, \qquad
g_s \rightarrow g_s\, R_s ,\label{eq:couplingrescale}
\end{equation}
where $\lambda$, $\beta$, and $g_s$ denote the baseline couplings appearing in the OBE potential, and the dimensionless rescaling factors $R_\lambda$, $R_\beta$, and $R_s$ encode the correlation between the interaction strength and the cutoff dependence. This parametrization effectively absorbs the dominant short-distance regulator effects into the couplings, while preserving the long-range structure dictated by meson exchange.

The rescaling factors are fixed by reproducing the pole structures of near-threshold exotic states in the two-body sector. The bound state poles corresponding to the $X(3872)$ and $T_{cc}(3875)$ are fixed at $0.4~\mathrm{MeV}$ below their respective thresholds—i.e., the $D\bar D^{*}$ threshold for the $X(3872)$ and the $DD^{*}$ threshold for the $T_{cc}(3875)$—as their exact positions lead to only marginal variations in the extracted coupling constants. In contrast, the virtual state pole position of the $Z_c(3900)$ state, which is associated with the $D\bar D^{*}$ threshold in the $C=-1$ channel (with $C$ denoting the charge-conjugation parity of the neutral $D\bar D^{*}$ system), exhibits a much stronger sensitivity to the interaction strength. To quantify this effect and propagate the corresponding uncertainty into the three-body sector, the $Z_c(3900)$ pole is varied from $5$ to $35~\mathrm{MeV}$ below the $D\bar D^{*}$ threshold. For each cutoff value of $\Lambda$, the corresponding sets of $(R_\lambda, R_\beta, R_s)$ are determined. These coupling constants are then employed consistently in the calculations of the $D^{(*)}NN$ systems, without introducing additional free parameters.

\subsection{Gaussian Expansion Method and Complex Scaling Method} \label{sec:method}

With the Hamiltonian specified above, the three-body Schr\"odinger equation for the $D^{(*)}NN$ system 
in the center-of-mass (cms) frame reads
\begin{equation}
H \Psi_{JM} = E \Psi_{JM},
\end{equation}
where $\Psi_{JM}$ denotes the total wave function with good total angular momentum $J$ and projection $M$.

\begin{figure}
  \centering
  \includegraphics[width=0.5\textwidth]{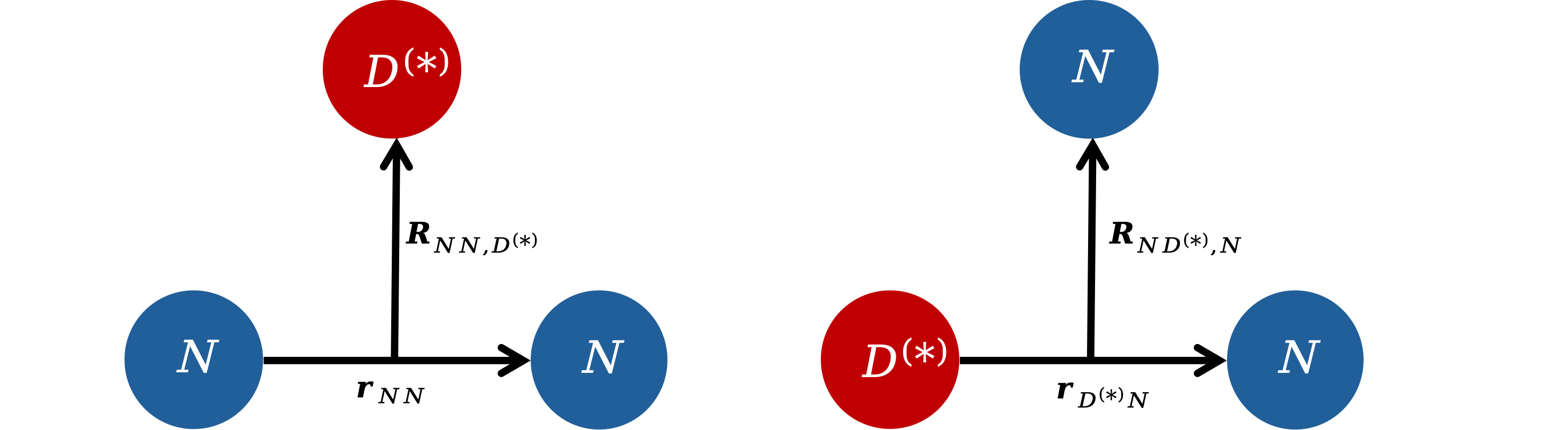}
  \caption{Two sets of Jacobi coordinates corresponding to different spatial
  configurations, where blue disks indicate the nucleons $N$ and the red disk
  denotes the $D^{(\ast)}$ meson. The third set of Jacobi coordinates, obtained by
  exchanging the two nucleons in panel (b), is not shown explicitly, but is
  included in the calculation.}
  \label{fig:jacobi}
\end{figure}

In the framework of GEM, the wave function is expanded in Gaussian bases defined on Jacobi coordinates,
\begin{equation}
\Psi_{JM} = \sum_{c} \sum_{n_1 n_2} 
C^{(c)}_{n_1 n_2} 
\Bigl[
\phi_{n_1 l_1}(\boldsymbol{r}_c) \phi_{n_2 l_2}(\boldsymbol{R}_c) 
\otimes \chi_s
\Bigr]_{JM}\, ,
\end{equation}
where $c$ labels different rearrangement channels, and $\chi_s$ is the spin wave function. The two sets of Jacobi coordinates corresponding to different rearrangement channels are illustrated in Fig.~\ref{fig:jacobi}. The spatial basis functions are
\begin{equation}
\phi_{nlm}(\boldsymbol{r}) = N_{nl} \, r^l e^{-r^2/r_n^2} Y_{lm}(\hat{\boldsymbol{r}}),
\quad r_n = r_1 a^{n-1}\, ,
\end{equation}
with Gaussian ranges in geometric progression, which efficiently describe both 
short-range correlations and long-range asymptotics. In the present calculation, the Gaussian range parameters are chosen as $r_{\min}=0.1~\mathrm{fm}$ and $r_{\max}=10~\mathrm{fm}$, with the number of basis functions set to $n_{\max}=12$ for each relative coordinate. The expansion coefficients $C^{(c)}_{n_1 n_2}$ are determined variationally via the Rayleigh--Ritz method.

To study resonances, we employ CSM, in which
coordinates and momenta are transformed as
\begin{equation}
\boldsymbol{r} \to \boldsymbol{r} e^{i\theta},  
\boldsymbol{p} \to \boldsymbol{p} e^{-i\theta}\, ,
\end{equation}
leading to the complex-scaled Hamiltonian $H(\theta)$. Bound states and
resonances appear as isolated eigenvalues $E = E_R - i \Gamma/2$ that are stable
under variations of $\theta$, while the continuum rotates by $2\theta$ in the
complex-energy plane. The Gaussian basis allows diagonalization of $H(\theta)$
in the same variational space as the bound-state calculation.

The root-mean-square (RMS) radius serves as a significant physical quantity for
characterizing the spatial structure of the studied system. Moreover, it is
widely utilized to determine whether a multi-quark system corresponds to a
molecular state or a compact one~\cite{Wu:2024zbx,Wu:2024euj}. For a bound state, the RMS
radius is defined as
\begin{equation}
r_{ij}^{\mathrm{rms}} \equiv 
\sqrt{
\frac{
\left\langle \Psi^{IJ} \middle| r_{ij}^{2} \middle| \Psi^{IJ} \right\rangle
}{
\left\langle \Psi^{IJ} \middle| \Psi^{IJ} \right\rangle
}
}.
\label{eq:rms_bound}
\end{equation}

For a resonant state within CSM, one can introduce
the RMS radius as
\begin{equation}
r_{ij}^{\mathrm{rms}} \equiv 
\mathrm{Re}
\left[
\sqrt{
\frac{
\left\langle \Psi^{IJ}(\theta) \middle| r_{ij}^{2} e^{2 i \theta} \middle| \Psi^{IJ}(\theta) \right\rangle
}{
\left\langle \Psi^{IJ}(\theta) \middle| \Psi^{IJ}(\theta) \right\rangle
}
}
\right],
\label{eq:rms_csm}
\end{equation}
where the complex-scaled product (c-product) is introduced in the following form to ensure
the analyticity of the integral~\cite{ROMO1968617}:
\begin{equation}
\left\langle \Psi_i \middle| \Psi_j \right\rangle
\equiv
\int \Psi_i(\boldsymbol{r}) \, \Psi_j(\boldsymbol{r}) \, d^{3}\boldsymbol{r}.
\label{eq:c_product}
\end{equation}

%%%%%%%%%%%%%%%%%%%%%%%%%%%%%%%%%%%%%%%%%%%%%%%%%%%%%%%
\section{Numerical Results and Analysis}\label{results}

Based on the established two-body framework, we extend the same interaction parameters and numerical methods to the three-body $D^{(*)}NN$ systems with $S$-wave coupling. In the two-body $NN$ interaction, due to the Pauli Exclusion Principle, only the deuteron channel with spin-orbital notation ${}^{2S+1}L_J = {}^3S_{1}$ and isospin $I_{NN}=0$ exists in the $S$-wave. Following the model and parameters adopted in Ref.~\cite{Machleidt:2000ge}, our calculation with a cutoff $\tilde{\Lambda} =0.814\ \text{GeV}$ indeed yields a deuteron-like bound state with a binding energy of $2.23\ \text{MeV}$ and $R_{rms}$ =3.42 fm. In the subsequent three-body calculations, we maintain exactly the same interaction parameters.

In the $DNN$ system, since the $D$ meson has spin $0$ and isospin $1/2$, and it couples in the $S$-wave to the $NN$ subsystem $I(J^P)=0(1^+)$, the total isospin of the system is fixed as $I=1/2$, and the total spin is $J=1$. For the $D^*NN$ system, the $D^{*}$ meson has spin $1$ and isospin $1/2$. With the total isospin constrained to $I=1/2$, the total spin $J$ can take the values $0$, $1$, or $2$ via angular momentum coupling. In the results presented below, we will analyze and discuss these three spin configurations separately.

With the two-body inputs and the quantum-number assignments of the three-body systems specified, we now construct the $D^{(*)}NN$ interaction. 
In the present work, the interaction is described within OBE framework, where the long- and intermediate-range dynamics are generated by light-meson exchanges, while the short-distance behavior is regulated by a cutoff parameter $\Lambda$. 
Such a regulator is inherent to effective hadronic descriptions and reflects unresolved short-range physics beyond the explicit degrees of freedom included in the model. Consequently, the resulting observables may retain a residual dependence on $\Lambda$ in the absence of additional short-range counterterms.

\begin{table*}[ht]
\caption{Comparison of the binding energies (in MeV) for the $S$-wave $DN$ and $D^{*}N$ systems with isospin $I$ and total spin $S$,
obtained in different approaches. 
The results of this work correspond to the cutoff $\Lambda = 1.2~\mathrm{GeV}$. The RMS radii are given in units of fm. Here ``ub" denotes that the corresponding channel is unbound. In this work, numerical results spanning a range mainly due to couplings constrained by the $Z_c(3900)$ virtual-state pole variation.}
\label{tab:DN_DsN_comparison}
\begin{ruledtabular}
\begin{tabular}{c c c c c c c}
System 
& $I(J^{P})$ 
& ChEFT \cite{Wang:2020dhf}
& ChQM \cite{Entem:2016lzh}
& QDCSM \cite{Zhao:2016zhf}
& This work 
&$R_{rms}$ \\
\hline
$DN$ 
& $0(1/2^-)$ 
& $11.1$ 
& 1.70
& --
& $0.35$ -- $4.54$ 
& $2.23$ -- $5.59$ \\

& $1(1/2^-)$ 
& -- 
& --
& -- 
& ub -- $2.72$ 
& ub -- $5.79$ \\

$D^{*}N$ 
& $0(1/2^-)$ 
& $1.5$ 
& --
& --
& -- 
& -- \\

& $1(1/2^-)$ 
& -- 
& 0.48 
& -- 
& $0.82$ -- $25.84$ 
& $1.06$ -- $4.36$ \\

& $0(3/2^-)$ 
& $6.7$ 
& 8.02
& 3.4 
& $8.35$ -- $43.23$ 
& $0.88$ -- $1.70$ \\

& $1(3/2^-)$ 
& -- 
& --
& -- 
& ub -- $0.33$
& ub -- $5.76$\\
\end{tabular}
\end{ruledtabular}
\end{table*}
%%%%%%%%%%%%%%%%%%%%%%%%%%%%%%%%%%%%%%%%%%%%%%%%%%%%%%%%%%%%%%%%%%%%%%
\subsection{$D^{(*)}N$ system: binding energies and spatial structure}

Table~\ref{tab:DN_DsN_comparison} presents a systematic comparison of the binding energies of the $S$-wave $DN$ and $D^{*}N$ systems with different isospin $I$ and total spin $S$ obtained in various theoretical approaches. The results of the present work are shown together with those from chiral effective field theory (ChEFT)~\cite{Wang:2020dhf}, chiral quark models (ChQM)~\cite{Entem:2016lzh}, and the quark delocalization color screening model (QDCSM)~\cite{Zhao:2016zhf}, providing a common basis for comparing the near-threshold behavior predicted by different theoretical approaches. 

For the $DN$ system, the isoscalar channel with $I=0$ and spin-parity $J^P=\tfrac{1}{2}^-$ is found to be attractive in most approaches, although the predicted binding energies vary significantly among different models. ChEFT predicts a relatively stronger attraction in this channel~\cite{Wang:2020dhf}, while quark model-based studies generally obtain either a shallow bound state or no binding at all unless additional dynamical mechanisms, such as channel coupling, are taken into account~\cite{Zhao:2016zhf}. In the present work, the isoscalar $DN$ channel supports only a very shallow bound state, accompanied by a relatively large RMS radius, indicating a near-threshold configuration. This behavior underscores the strong model dependence of the $DN$ interaction and its sensitivity to the treatment of short-range dynamics.

However, whether the isovector $DN$ channel with $I(J^P)=1(\tfrac{1}{2}^-)$ remains unbound is inconclusive in the present calculation. When adopting different $DD\sigma$ couplings constrained by the variation of the $Z_c(3900)$ virtual-state pole, the results shift between bound and unbound states. Nevertheless, even in the unbound scenarios, the interaction extracted in this work for this channel is attractive.  In the literature, no bound state is found in this channel in ChEFT~\cite{Wang:2020dhf} or QDCSM-based studies~\cite{Zhao:2016zhf}. Notably, however, some quark-model calculations predict the existence of a bound state~\cite{Entem:2016lzh,Zhang:2019vqe}, as also indicated in Refs.~\cite{Sakai:2020psu,Haidenbauer:2010ch}. 

A more robust and consistent pattern emerges for the $D^{*}N$ system. For the isoscalar channels, different theoretical approaches generally agree on the existence of attractive interactions. In particular, the $D^{*}N$ state with $I=0$ and $J^P=\tfrac{3}{2}^-$ is predicted to be bound in ChEFT~\cite{Wang:2020dhf}, ChQM calculations~\cite{Entem:2016lzh}, and QDCSM-based analyses~\cite{Zhao:2016zhf}. In the present work, this channel also supports a bound-state solution with a relatively small RMS radius, suggesting a compact molecular configuration. The qualitative agreement among different approaches highlights the enhanced stability of this channel and a reduced sensitivity to model-dependent short-range dynamics.

The $D^{*}N$ interaction has also been investigated within the one-boson-exchange (OBE) framework~\cite{He:2010zq,Haidenbauer:2010ch}. While the present results are qualitatively consistent with these OBE studies, differences in the cutoff dependence are observed. In particular, bound-state solutions in certain $D^{*}N$ channels appear only when the cutoff parameter exceeds a moderate value, indicating that the detailed quantitative predictions remain sensitive to the regularization scheme. Such differences suggest that the discrepancies among various OBE-based calculations originate primarily from the treatment of short-range regularization.

\begin{figure*}[ht]
\centering
\includegraphics[width=0.98\textwidth]{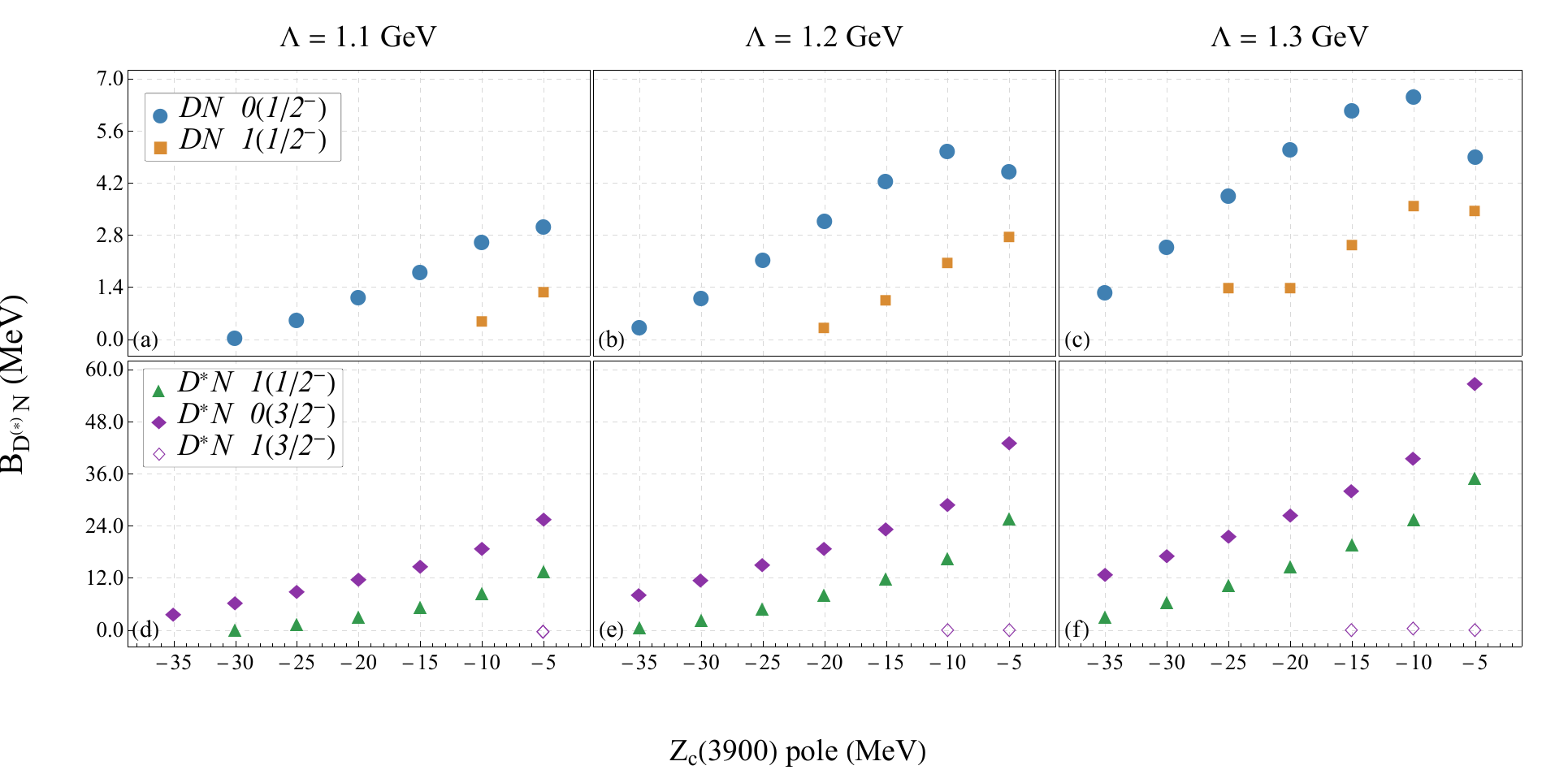}
% includegraphics here
\caption{
Binding energies of the $S$-wave $DN$ and $D^{*}N$ systems 
as functions of the $Z_c(3900)$ virtual-state pole position, 
for three representative cutoff values 
$\Lambda=1.1$, $1.2$, and $1.3~\mathrm{GeV}$. 
Each column corresponds to a fixed cutoff. 
The overall trend illustrates how the two-body binding energies 
evolve with the input pole position used to constrain the interaction.
}
\label{fig:DN_DstarN_BE}
\end{figure*}

\begin{figure*}[ht]
\centering
\includegraphics[width=0.98\textwidth]{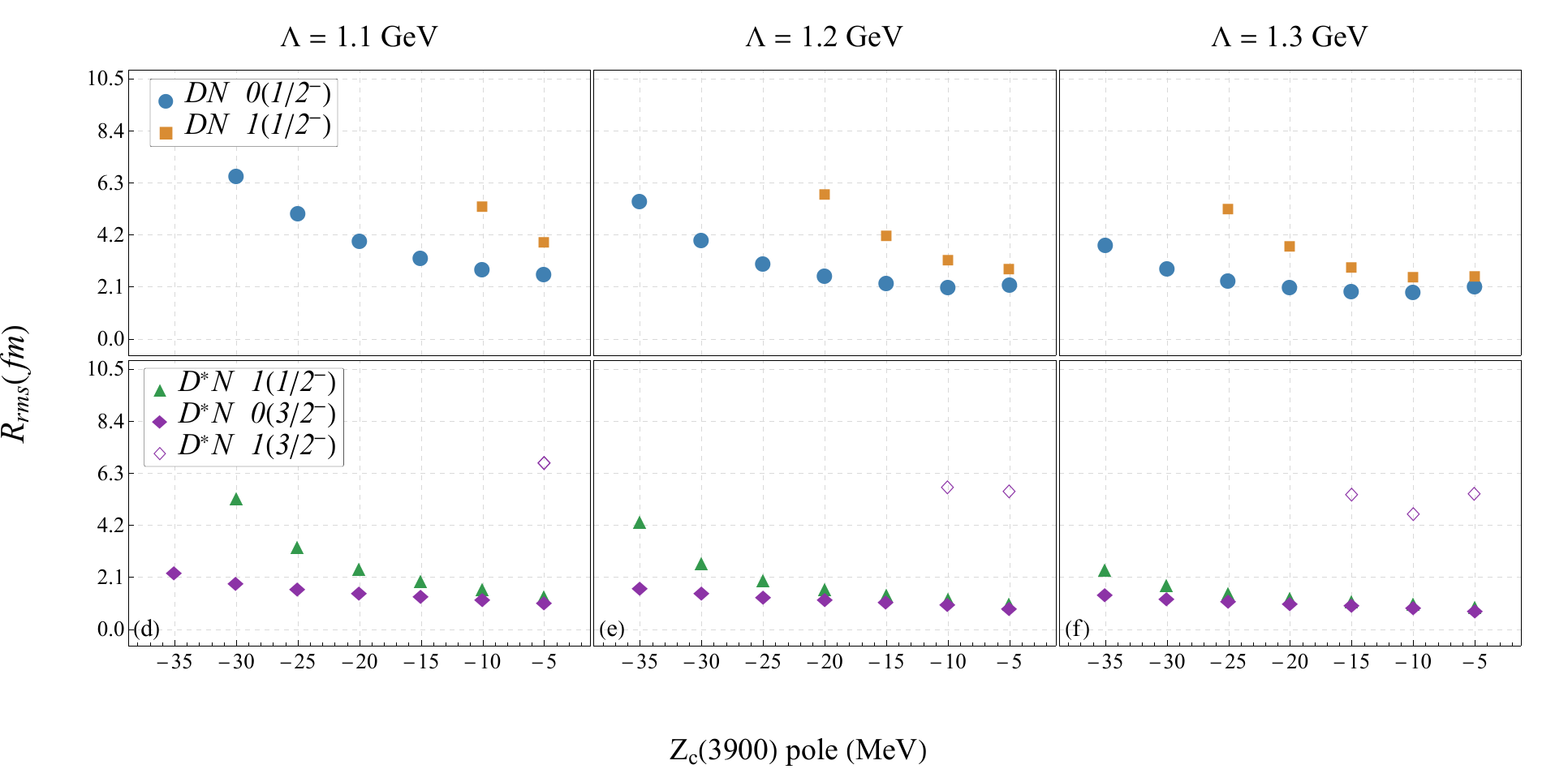}
\caption{
Root-mean-square (RMS) radii of the $S$-wave $DN$ and $D^{*}N$ systems 
as functions of the $Z_c(3900)$ virtual-state pole position, 
for the same cutoff values as in Fig.~\ref{fig:DN_DstarN_BE}. 
The spatial sizes increase as the binding energies decrease, 
indicating near-threshold and spatially extended configurations.
}
\label{fig:DN_DstarN_RMS}
\end{figure*}

To further illustrate the regulator and input dependence of the two-body sector, 
we show in Figs.~\ref{fig:DN_DstarN_BE} and~\ref{fig:DN_DstarN_RMS} the binding energies and RMS radii of the $S$-wave $DN$ and $D^{*}N$ systems as functions of the $Z_c(3900)$ virtual-state pole position, for three representative cutoffs $\Lambda=1.1$, $1.2$, and $1.3~\mathrm{GeV}$. As the $Z_c(3900)$ pole is moved away from the threshold (from $-5$ to $-35~\mathrm{MeV}$), the $DD\sigma$ couplings become weaker, leading to systematically $D^{(*)}N$ reduced binding energies and increased spatial extents in all channels that exhibit attraction.
This trend provides a transparent diagnostic of how the two-body inputs propagate into the three-body calculations. For the isoscalar $DN$ channel with $I(J^P)=0(\tfrac{1}{2}^-)$, the interaction remains only weakly attractive in the explored parameter region. A shallow bound state can appear for moderate-to-large cutoffs and for relatively shallow $Z_c$ poles, while it quickly dissolves when the couplings are weakened by shifting the $Z_c$ pole to more negative energies. 
Correspondingly, the RMS radius increases rapidly in the near-threshold region, indicating a spatially extended configuration characteristic of a weakly bound state. In contrast, the isovector $DN$ channel $I(J^P)=1(\tfrac{1}{2}^-)$ does not support a bound-state solution for the baseline parameters in the entire range shown in the figures, although the overall trend indicates that the interaction is still mildly attractive, consistent with the discussion above. A more robust pattern is observed in the $D^{*}N$ sector. 
The isoscalar channel with $I(J^P)=0(\tfrac{3}{2}^-)$ exhibits the strongest and most stable binding among the channels considered: it remains bound throughout a broad range of $Z_c$ pole positions and cutoffs, and its RMS radius stays comparatively small, pointing to a compact molecular configuration. By contrast, the isovector $D^{*}N$ channel with $I(J^P)=1(\tfrac{1}{2}^-)$ is much more sensitive to the short-range regularization and the fitted couplings: it typically appears as a very shallow bound state only for sufficiently strong couplings (i.e., for a shallow $Z_c$ pole) and/or larger $\Lambda$, and is accompanied by large RMS radii, suggesting a near-threshold and spatially diffuse structure. No bound-state solution is found for the remaining $I=1$ $D^{*}N$ channels in the parameter region displayed. These two-body trends provide the baseline for the three-body analysis below. In particular, the coexistence of a robustly bound $D^{*}N$ isoscalar channel and several near-threshold (or unbound but attractive) channels implies that three-body correlations can substantially enhance binding and induce spatial compression even when individual two-body subsystems are only weakly bound.

Overall, the comparison summarized in Table~\ref{tab:DN_DsN_comparison} indicates that the isoscalar $D^{*}N$ channel, especially with $J^P=\tfrac{3}{2}^-$, exhibits the highest degree of stability across different theoretical approaches and thus represents the most promising molecular candidate. In contrast, the binding properties of the $DN$ system are found to be significantly more model-dependent and sensitive to the details of the short-range interaction.

%%%%%%%%%%%%%%%%%%%%%%%%%%%%%%%%%%%%%%%%%%%%%%%%%%%%%%%%%%%%%%%%%%%%%%%
\subsection{$D^{(*)}NN$ system: binding energies and spatial structure}

\begin{figure*}[t]
\centering

\begin{overpic}[width=0.98\textwidth]{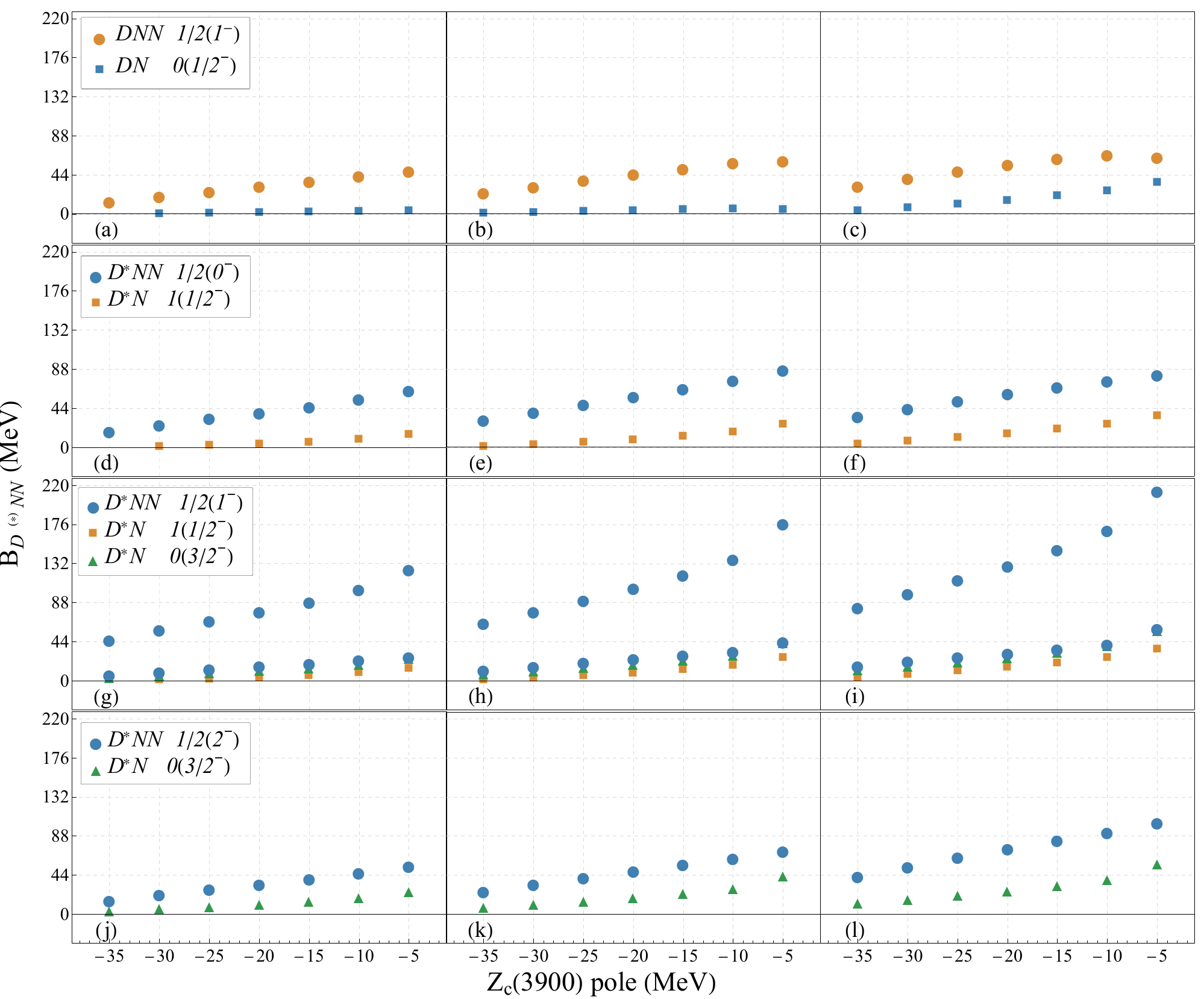}
\put(16,83){\fontsize{11}{20}\selectfont $\Lambda = 1.1\,\mathrm{GeV}$}
\put(47,83){\fontsize{11}{20}\selectfont $\Lambda = 1.2\,\mathrm{GeV}$}
\put(78,83){\fontsize{11}{20}\selectfont $\Lambda = 1.3\,\mathrm{GeV}$}
\end{overpic}

\caption{
Binding energies of the three-body $D^{(*)}NN$ and the corresponding two-body subsystems, calculated using coupling constants fitted to the $Z_c(3900)$ pole, are shown as functions of the $Z_c(3900)$ virtual state pole position (from $-5$ to $-35$ MeV). The panels are organized in columns of fixed $\Lambda$ (1.1, 1.2, 1.3 GeV) and rows corresponding to specific physical systems: three spin configurations of $D^*NN$ and one $DNN$ system. Each panel displays both three-body $D^{(*)}NN$ and two-body $D^{(*)}N$ binding energies, enabling a direct comparison of cutoff dependence, sensitivity to the $Z_c(3900)$ pole, and the interplay between two-
and three-body dynamics.}

\label{fig:binding}

\end{figure*}

\begin{figure*}[t]
\centering
\includegraphics[width=0.98\textwidth]{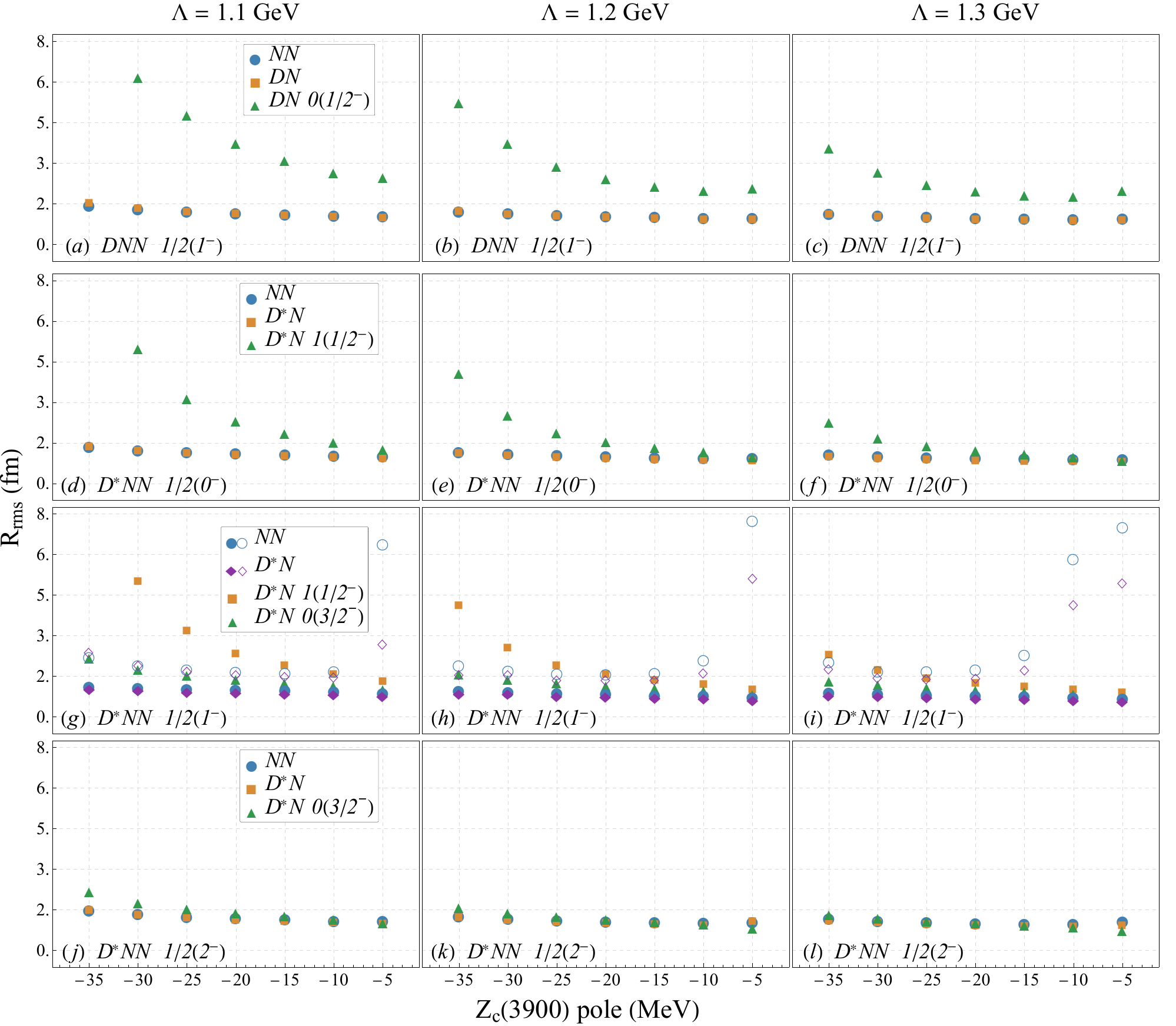}
\caption{
RMS radii of the $D^{(*)}NN$ systems and the corresponding two-body subsystems, calculated with the same coupling constants $(\lambda,\beta, g_s)$ as in Fig.~\ref{fig:binding}, which are determined by fitting the $Z_c(3900)$ pole. The RMS radii are shown as functions of the $Z_c(3900)$ pole position from $-5$ to $-35~\mathrm{MeV}$ for three cutoff values $\Lambda=1.1$, $1.2$, and $1.3~\mathrm{GeV}$. The panel layout is identical to that of Fig.~\ref{fig:binding}: each column corresponds to a fixed cutoff $\Lambda$, and each row to a specific hadronic system. In each panel, the RMS radii of the three-body $D^{(*)}NN$ states are shown together with those of the corresponding two-body $D^{(*)}N$ subsystems, allowing for a direct comparison of their spatial structures and their dependence on the cutoff and the $Z_c(3900)$ pole.
}
\label{fig:rms}
\end{figure*}

The resulting binding energies and RMS radii of the $D^{(*)}NN$ systems are summarized in Figs.~\ref{fig:binding} and~\ref{fig:rms}, respectively. %For a transparent comparison with the corresponding three-body states, we display in each partial-wave channel the most deeply bound two-body state among the $D^{(*)}N$ subsystems as a reference. We first discuss the $DNN$ system. 
To figure out the nature of the three-body states, we additionally display, in each 
partial-wave channel, the most deeply bound (or closest-to-threshold)  two-body $D^{(*)}N$ 
state as a reference. By comparing the three-body energies with those of the 
relevant two-body systems, one can read whether the obtained states correspond 
to genuine three-body bound states as well as to understand the role of the third 
particle in the formation of the three-body configurations.
%\textcolor{red}{For a transparent comparison with the corresponding three-body states, we also display, in each partial-wave channel, the most deeply bound (or closest-to-threshold) two-body $D^{(*)}N$ state as a reference. 
%This allows one to directly compare the three-body energies with the relevant two-body thresholds and to assess whether the obtained states correspond to genuine three-body bound states as well as to understand the role of the two-body subsystems in the formation of the three-body configurations.}
For $\Lambda=1.1~\mathrm{GeV}$, the three-body binding energy decreases from $47.76~\mathrm{MeV}$ at a $Z_c$ pole of $-5~\mathrm{MeV}$ to $13.26~\mathrm{MeV}$ at $-35~\mathrm{MeV}$. A similar monotonic behavior is observed for $\Lambda=1.2$ and $1.3~\mathrm{GeV}$, where the binding energies span the ranges $59.71$-$23.17~\mathrm{MeV}$ and $63.62$-$31.03~\mathrm{MeV}$, respectively. These values are substantially larger than the binding energy of the two-body $NN$ system alone, $B_{NN}=2.2~\mathrm{MeV}$.

This comparison already demonstrates that the $DNN$ bound states are not simply perturbations of the deuteron-like $NN$ subsystem. Instead, the additional attraction provided by the $DN$ interaction plays an essential role in stabilizing the three-body configuration. This conclusion is reinforced by the analysis of the spatial structure. As shown in Fig.~\ref{fig:rms}, the RMS radii of the $DNN$ system are typically of order $1~\mathrm{fm}$. For example, at $\Lambda=1.2~\mathrm{GeV}$ the RMS radius of the dominant $NN$ component varies from $1.04~\mathrm{fm}$ to $1.31~\mathrm{fm}$ as the $Z_c$ pole is shifted. These values are more than a factor of three smaller than the RMS radius of the two-body $NN$ bound state, which is approximately $3.4~\mathrm{fm}$. The pronounced spatial compression indicates that the $D$ meson acts as an effective attractor that draws the two nucleons into a compact configuration, rather than merely orbiting a pre-existing $NN$ core.

We next turn to the $D^{*}NN$ system with total spin $0^-$. In this channel, the binding energy is systematically larger than that for the  $DNN$ system. For instance, at $\Lambda=1.2~\mathrm{GeV}$ the binding energy decreases from $86.95~\mathrm{MeV}$ to $30.28~\mathrm{MeV}$ as the $Z_c$ pole is moved from $-5$ to $-35~\mathrm{MeV}$. The corresponding RMS radii of the $NN$-dominated component remain in the range $1.0$--$1.3~\mathrm{fm}$, indicating compact three-body configurations. At the same time, the RMS radii associated with the $D^{*}N$ coordinates increase rapidly as the binding weakens, reaching values of several femtometers in the shallow-binding region. This behavior reflects the marginal binding of the $D^{*}N$ two-body subsystem and highlights the nontrivial role of three-body correlations: even when one of the two-body subsystems is weakly bound or unbound, the three-body system can still form a deeply bound and compact state.

A particularly rich structure is found in the $D^{*}NN$ system with total spin $S=1$, where two distinct branches emerge. The upper branch corresponds to deeply bound states dominated by short-range dynamics. For $\Lambda=1.1~\mathrm{GeV}$, the binding energy of the upper branch reaches $124.75~\mathrm{MeV}$ at the shallowest $Z_c$ pole, while for $\Lambda=1.3~\mathrm{GeV}$ it exceeds $200~\mathrm{MeV}$. The RMS radii of these states are typically below $1~\mathrm{fm}$. For example, at $\Lambda=1.3~\mathrm{GeV}$ the RMS radius of the $NN$ component in the upper branch varies from $0.71~\mathrm{fm}$ to $0.96~\mathrm{fm}$. Such small spatial extensions are characteristic of strongly correlated, short-range dominated systems and indicate that these states probe the deep interior of the effective interaction.

In contrast, the lower $1^-$ branch exhibits a qualitatively different behavior. The binding energies are more moderate, ranging from a few MeV to several tens of MeV, and the spatial structure shows a pronounced sensitivity to the $Z_c$ pole position and the cutoff. For $\Lambda=1.1~\mathrm{GeV}$ and a $Z_c$ pole at $-5~\mathrm{MeV}$, the RMS radius of the $NN$ component reaches $6.8~\mathrm{fm}$, which is comparable to or even larger than the typical size of weakly bound nuclear systems. Such large spatial extensions signal near-threshold, halo-like configurations. As the $Z_c$ pole moves deeper or as the cutoff increases, these lower branch states become progressively more compact, with RMS radii decreasing toward the $1.5$-$2~\mathrm{fm}$ range. This strong evolution illustrates a delicate balance between short-distance attraction and long-range dynamics.

Finally, we consider the $D^{*}NN$ system with total spin $2^-$. In this channel, a single bound branch is found. The binding energies range from approximately $15~\mathrm{MeV}$ to nearly $100~\mathrm{MeV}$, depending on $\Lambda$ and the $Z_c$ pole position. The RMS radii again exhibit a clear anticorrelation with the binding energy, varying from about $1.1~\mathrm{fm}$ for deeply bound states to approximately $1.6~\mathrm{fm}$ near threshold. Compared with the deeply bound $1^-$ upper branch, the $2^-$ states are less compact, yet they remain significantly more localized than the corresponding two-body $D^{*}N$ subsystems.

In summary, the combined analysis of binding energies and RMS radii reveals a rich spectrum of structures for the $D^{(*)}NN$ systems. Depending on the spin configuration and the two-body input, the systems can manifest as deeply bound and compact states or as weakly bound, spatially extended configurations. While the two-body $NN$ and $D^{(*)}N$ subsystems are %either weakly bound or extended
weakly bound (normally with spatially-extended configuration), three-body correlations can generate substantially stronger
binding and pronounced spatial compression. The residual cutoff dependence, particularly visible in the deeply bound $1^-$ states, reflects the importance of short-range physics and provides an estimate of the theoretical uncertainty associated with the absence of
an explicit three-body counter term.

\begin{figure*}[t]
\centering
\begin{minipage}[b]{0.48\textwidth}
  \centering
  \includegraphics[width=\linewidth]{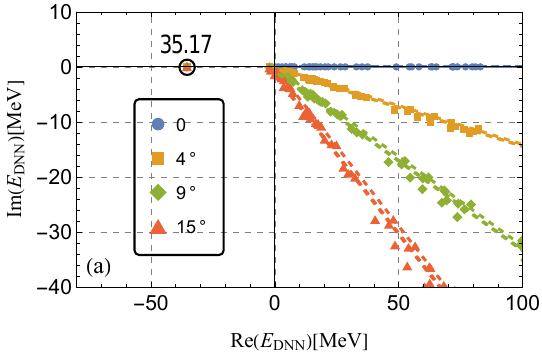}
  \label{fig:pole_a}
\end{minipage}\hfill
\begin{minipage}[b]{0.48\textwidth}
  \centering
  \includegraphics[width=\linewidth]{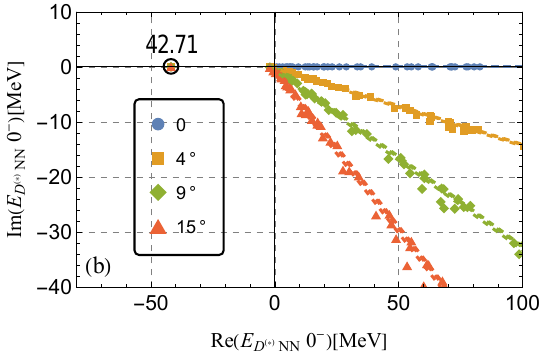}
  \label{fig:pole_b}
\end{minipage}

\vspace{2mm}

\begin{minipage}[b]{0.48\textwidth}
  \centering
  \includegraphics[width=\linewidth]{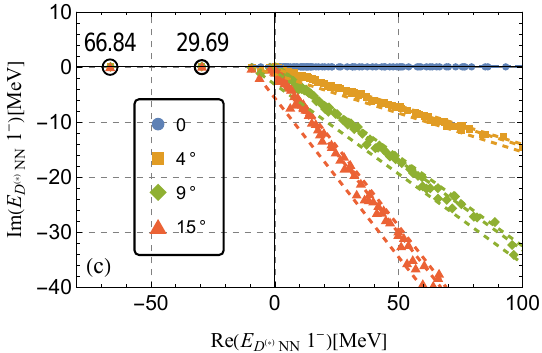}
  \label{fig:pole_c}
\end{minipage}\hfill
\begin{minipage}[b]{0.48\textwidth}
  \centering
  \includegraphics[width=\linewidth]{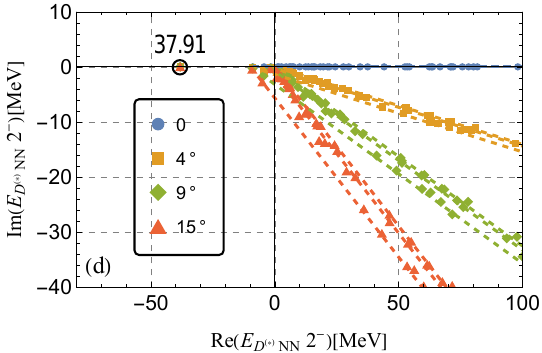}
  \label{fig:pole_d}
\end{minipage}

\caption{Pole distributions of the $D^{(*)}NN$ systems in the complex energy plane for (a) $DNN$, (b) $D^{*}NN$ with $J^P=0^-$, (c) $D^{*}NN$ with $J^P=1^-$,and (d) $D^{*}NN$ with $J^P=2^-$. The calculations are performed with $R_s=R_\lambda=R_\beta=1$ and a fixed cutoff $\Lambda=1.4~\mathrm{GeV}$. The real and imaginary parts of the three-body energies are shown as the interaction strength is varied. In each panel, the trajectories originate from the three-body threshold, the $D^{(*)}N$ two-body threshold, and the $NN$ two-body threshold, respectively. The poles evolve along the real-energy axis indicate genuine bound-state poles, while the absence of poles with sizable imaginary parts confirms the non-resonant nature of the states. In the $D^{*}NN$ $J^P=1^-$ channel, two distinct bound-state poles are
observed, corresponding to the two branches identified in the binding energy and rms analyses.}
\label{fig:pole}
\end{figure*}

% To further clarify the nature of the bound states obtained
% in the present
% calculation, we additionally examine the pole trajectories of the
% three-body $D^{(*)}NN$ systems in the complex energy plane.
% The results are shown in Fig.~\ref{fig:pole}, where the
% real and imaginary parts of the three-body energies are displayed for
% the $DNN$ and $D^{*}NN$ systems with $J^P=0^-$, $1^-$, and $2^-$.
% We note that the qualitative features discussed below are not
% sensitive to the specific choice of the rescaling factors.
% For simplicity and clarity of presentation, we set
% $R_s=R_\lambda=R_\beta=1$ and adopt a fixed cutoff
% $\Lambda=1.4~\mathrm{GeV}$ in Fig.~\ref{fig:pole}.The corresponding numerical results are summarized in
% Tables.~\ref{tab:DNN_results} and \ref{tab:DstarNN_results} in the Appendix.

To further clarify the nature of the bound states obtained in the present calculation, 
we additionally examine the pole distribution of the three-body $D^{(*)}NN$ systems in the complex 
energy plane. The results are shown in Fig.~\ref{fig:pole}, where the real and imaginary parts of the 
three-body energies are displayed for the $DNN$ and $D^{*}NN$ systems with $J^P=0^-$, $1^-$, and $2^-$.
Here we adopt the conventional ($\Lambda$-independent) scheme, 
in which the coupling constants are fixed and no rescaling factors are introduced. 
In this framework, the results are independent of any rescaling parameters, allowing for a more 
direct comparison with previous studies.
For simplicity and clarity of presentation, we therefore set
$R_s = R_\lambda = R_\beta = 1$
and use a fixed cutoff $\Lambda = 1.4~\mathrm{GeV}$ in Fig.~\ref{fig:pole}.
We emphasize that this choice does not imply any parameter tuning, but  rather, 
it demonstrates that the obtained bound-state structures persist within the conventional 
parameter setup. The  numerical results for different cutoffs $\Lambda = 1.35, 1.40$, and $1.45$ GeV, 
reflecting the dependence on the short-range regularization are summarized in Tables~\ref{tab:DNN_results} 
and \ref{tab:DstarNN_results} in the Appendix.

%\textcolor{red}{Old:We note that the qualitative features discussed below are not sensitive to the specific choice of the rescaling factors; for simplicity and clarity of presentation, all rescaling factors are set to unity in Fig.~\ref{fig:pole}.
%These calculations are performed within the rescaling scheme with $R_s=R_\lambda=R_\beta=1$ and a fixed cutoff $\Lambda=1.4~\mathrm{GeV}$.}

In each panel, three reference thresholds are explicitly indicated,
corresponding to the three-body threshold, the two-body $D^{(*)}N$
threshold, and the two-body $NN$ threshold. The pole trajectories originate from these thresholds as the interaction strength is gradually increased, forming three distinct rays in the complex energy plane. This representation provides a direct and unambiguous diagnostic of the analytic structure of the three-body spectrum.

A salient feature of Fig.~\ref{fig:pole} is that all poles
associated with the $DNN$ and $D^{*}NN$ systems move strictly along the
negative real-energy axis without developing a sizable imaginary part.
This behavior demonstrates that the states obtained in the present work
correspond to genuine bound states rather than resonant or virtual
states. In particular, no pole is observed to cross into the lower-half complex plane away from the real axis, which would be characteristic of a three-body resonance. The absence of such behavior provides strong evidence that the deeply bound states found in Figs.~\ref{fig:binding} and~\ref{fig:rms} are not artifacts of threshold effects or numerical continuation, but represent true bound solutions of the three-body problem.

For the $DNN$ and $D^{*}NN$ systems with $J^P=0^-$ and $2^-$, a single
dominant bound-state pole is observed.
The corresponding trajectories originate from the three-body threshold
and rapidly move to large negative real energies, indicating deeply
bound configurations.
The fact that these poles remain well separated from both the $D^{(*)}N$
and $NN$ thresholds further confirms their compact and strongly bound
nature.

In contrast, the $D^{*}NN$ system with $J^P=1^-$ exhibits 
a qualitatively different behavior.
As shown in Fig.~\ref{fig:pole}(c), two distinct bound-state
poles are clearly identified.
One pole originates from the three-body threshold and rapidly evolves
into a deeply bound state, while the second pole emerges closer to the
two-body thresholds and remains comparatively shallow.
This two-pole structure is consistent with the two-branch pattern
observed in the binding-energy and RMS-radius analyses, and reflects the
interplay between different spin-coupling mechanisms and short-range
dynamics of $D^{*}NN$ in the $J^P = 1^-$ channel.

Taken together, the pole-distribution analysis provides a stringent
confirmation of the bound-state nature of the $D^{(*)}NN$ systems
identified in this work. The simultaneous absence of resonance poles and the presence of multiple bound-state poles in the $D^{*}NN$ $1^-$ channel strongly support the physical interpretation of these states as genuine three-body bound states generated by the underlying two-body interactions and three-body correlations.

%%%%%%%%%%%%%%%%%%%%%%%%%%%%%%%%%%%%%%
\section{Conclusion}\label{conclusion}

In this work, we have carried out a systematic and unified investigation of the $DNN$ and $D^{*}NN$ three-body systems within a hadronic molecular
framework. Motivated by the growing experimental and theoretical interest in heavy-meson-nucleon dynamics, and guided by the interplay of heavy-quark spin symmetry and realistic nuclear forces emphasized in the Introduction. We constructed the three-body Hamiltonian using a realistic $NN$ interaction together with an energy-independent $D^{(*)}N$ potential.
The latter is constrained by heavy-quark spin symmetry and incorporates
long- and intermediate-range dynamics generated by light-meson exchange.
The three-body Schr\"odinger equation was solved using GEM, while CSM was employed to analyze the analytic structure of the spectrum and to distinguish bound states from possible resonances.

At the two-body level, our analysis confirms that the 
isoscalar $D^{*}N$ channel with $J^P=3/2^-$ provides 
the most robust attraction across different
theoretical descriptions, whereas the binding 
properties of the $DN$ system are significantly 
more model-dependent and sensitive to short-range dynamics.
These features set the stage for the three-body problem.
%and highlight the
%importance of coupled-channel effects and tensor forces induced by the
%near-degeneracy of the $D$ and $D^*$ mesons.

Extending the same interaction consistently to the 
three-body sector, we find that three-body correlations 
can generate substantially stronger binding than
suggested by the two-body subsystems alone. In 
the $DNN$ system, a deep bound state is obtained 
in the $I(J^P)=\tfrac{1}{2}(1^-)$ channel over a wide range of
parameters, even when the corresponding $DN$ subsystems 
are weakly bound or unbound. The associated RMS radii 
indicate a pronounced spatial compression compared 
with the deuteron, demonstrating that the heavy meson 
acts as an effective attractor that drives the nucleons 
into a compact configuration, rather than merely 
forming a loosely bound meson-deuteron-like state.

% The $D^{*}NN$ systems exhibit an even richer structure
% due to the spin-1 nature of the $D^*$ meson. Deeply 
% bound and compact states are found in both $J^P=0^-$
% and $2^-$ channels, while the $J^P=1^-$ channel displays 
% a characteristic two-branch pattern. 
%\textcolor{red}{The $D^{*}NN$ systems exhibit an even richer structure due to the spin-1 nature of the $D^*$ meson. 
%This pronounced spin dependence originates mainly from the tensor components of the $D^{(*)}N$ interaction, in particular those induced by pion exchange, which strongly mix different partial waves and generate spin-dependent attraction in the three-body system. Deeply bound and compact states are found in both $J^P=0^-$ and $2^-$ channels, while the $J^P=1^-$ channel displays a characteristic two-branch pattern.}
%% The upper branch corresponds to strongly bound, short-range dominated configurations with very small RMS radii, 
%\textcolor{red}{The upper branch 
%corresponds to strongly bound, short-range dominated 
%configurations with very small RMS radii.
%We note that for binding energies of order $\sim 100~\mathrm{MeV}$ or larger, the system probes distances where the constituent hadrons may significantly overlap. In this regime, the interpretation in terms of loosely bound hadronic molecules becomes less reliable, and the states should be regarded as compact configurations driven by strong attraction rather than pure molecular states.}

The $D^{*}NN$ systems exhibit even richer structures due to the spin-1 nature of the $D^*$ meson. 
This pronounced spin dependence originates mainly from the tensor components of the $D^*N$ interaction, 
in particular those induced by the pion exchange, which strongly mix different partial waves and 
generate spin-dependent attraction in the three-body system. Deeply bound and compact states 
are found in both $J^P=0^-$ and $2^-$ channels, while the $J^P=1^-$ 
channel displays a characteristic two-branch pattern. The upper branch 
corresponds to strongly bound, short-range dominated 
configurations with very small RMS radii. We note that in some cases the binding energies are of 
order $\sim100$~MeV or even larger. In such cases, the system 
probes distance where the constituent hadrons may significantly overlap. We emphasize that in this regime, 
the interpretation in terms of loosely bound hadronic molecules becomes less reliable, and 
the states may be regarded as compact configurations driven by strong attraction rather than pure 
molecular states.
Whereas the shallower branch is more weakly bound and spatially extended, approaching a near-threshold, halo-like structure 
for certain parameter choices. This clear spin hierarchy
and the emergence of multiple branches provide a direct 
manifestation of the spin-dependent forces.

%and tensor interactions associated with the coupled
%D^{(*)}N$ dynamics, as anticipated from heavy-quark spin symmetry arguments
%discussed in the Introduction.

The bound-state nature of these configurations is further 
corroborated by the pole-distribution analyses in the 
complex energy plane. All identified poles associated 
with the $DNN$ and $D^{*}NN$ systems evolve along the real-energy
axis and do not acquire sizable imaginary parts, excluding a resonant
interpretation. In particular, the two-branch structure observed in the
$D^{*}NN$ $J^P=1^-$ channel is reflected the appearance of two distinct
bound-state poles, providing a consistent and unambiguous analytic
interpretation of the spectrum.

In summary, our results demonstrate that the interplay of realistic $NN$
correlations, the $D^{(*)}N$ interactions, and heavy-quark spin
symmetry can give rise to deeply bound and spatially compact heavy-flavor
three-body states. These findings extend the concept of meson-nuclear bound
systems into the charm sector.
%and complement analogous studies in the strange sector. 
Future improvements, such as the inclusion of explicit three-body forces 
or a systematic effective field theory treatment, will be essential for
reducing theoretical uncertainties. On the experimental side, the predicted
binding energies, spin hierarchy, and spatial characteristics provide 
concrete targets for future searches for heavy-meson-nuclear bound states 
at facilities such as LHC, J-PARC, and GSI-FAIR.

\begin{acknowledgements}
We are grateful to Feng-Kun Guo, Ming-Zhu Liu, Si-Qiang Luo, Tian-Wei Wu, and Hai-Peng Xie for useful discussions. This work is supported by the Special Funds for Theoretical Physics under the National Natural Science Foundation of China (Grant No. 12547105), the National Natural Science Foundation of China (Grant No. 12575153) and the Start-up Funds of Southeast University (Grant No. 4007022506).
\end{acknowledgements}

\section*{Data Availability}
The data that support the findings of this study are openly available~\cite{chen_2026_20337529}.

%\appendix
%\section{Numerical details of the $D^{(*)}NN$ and $D^{(*)}N$ system }

\appendix
\section{Numerical details of the $D^{(*)}NN$ and $D^{(*)}N$ system}

\begin{table*}[htbp]
\centering
\renewcommand{\arraystretch}{1.5}
\setlength{\tabcolsep}{6pt}
\caption{Numerical results for the three-body $DNN$ system and the corresponding two-body subsystems are obtained with the rescaled coupling constants fixed at $R_s = R_\beta = R_\lambda = 1$. The three-body binding energy difference is defined as
$\Delta B = B - \max\left(B_{DN}, B_{NN}\right)$ where $B$ denotes the three-body binding energy, and $B_{DN}$ and $B_{NN}$ are the binding energies of the two-body subsystems. }
\label{tab:DNN_results}
\begin{tabular}{c c cc cc c c c c}
\hline\hline
 &  & \multicolumn{4}{c}{$DN$ subsystems} &  &  &  &  \\
\cline{3-6}
System & $\Lambda$ (GeV)
& \multicolumn{2}{c}{$0(\tfrac{1}{2}^-)$}
& \multicolumn{2}{c}{$1(\tfrac{1}{2}^-)$}
& $B$ (MeV)
& $\Delta B$ (MeV)
& $R_{rms}(NN)$ (fm)
& $R_{rms}(DN)$ (fm) \\
\cline{3-6}
 &  & $B$ (MeV) & $R_{rms}$ (fm) & $B$ (MeV) & $R_{rms}$ (fm) &  &  &  &  \\
\hline
\multirow{3}{*}{$DNN$}
& 1.35 & 1.10 & 4.03 & -- & -- & 25.23 & 23.00 & 1.26 & 1.21 \\
& 1.40 & 2.22 & 2.79 & -- & -- & 35.17 & 32.94 & 1.16 & 1.08 \\
& 1.45 & 4.72 & 2.18 & -- & -- & 46.49 & 41.77 & 1.07 & 0.98 \\
\hline\hline
\end{tabular}
\end{table*}

\begin{table*}[htbp]
\renewcommand{\arraystretch}{1.5}
\setlength{\tabcolsep}{6pt}
\centering
\caption{Numerical results for the three-body $D^{*}NN$ system and the corresponding two-body subsystems are obtained with the rescaled coupling constants fixed at $R_s = R_\beta = R_\lambda = 1$. 
To quantify whether a genuine three-body bound state is formed, we define the binding energy difference as
$\Delta B = B - \max\left(B_{D^{*}N}, B_{NN}\right)$,where $B$ denotes the three-body binding energy, and the maximum is taken over all relevant two-body subsystems.A positive value of $\Delta B$ indicates that the three-body state lies below the deepest two-body threshold and therefore corresponds to a genuine three-body bound state.}
\label{tab:DstarNN_results}
\begin{tabular}{c c c cc cc cc cc c c c c}
\hline\hline
%----------- add the "D*N subsystems" super header -----------
 &  &  & \multicolumn{8}{c}{$D^{*}N$ subsystems} &  &  &  &  \\
\cline{4-11}
 &  &  &
\multicolumn{2}{c}{$0(\tfrac12^-)$} &
\multicolumn{2}{c}{$1(\tfrac12^-)$} &
\multicolumn{2}{c}{$0(\tfrac32^-)$} &
\multicolumn{2}{c}{$1(\tfrac32^-)$} &
\multirow{2}{*}{$B$} &
\multirow{2}{*}{$\Delta B$} &
\multirow{2}{*}{$R_{rms}(NN)$} &
\multirow{2}{*}{$R_{rms}$($D^{(*)}N$)} \\
\cline{4-11}
System & $J^P$ & $\Lambda$ (GeV) &
$B$ & $R_{rms}$ & $B$ & $R_{rms}$ & $B$ & $R_{rms}$ & $B$ & $R_{rms}$ &
 &  &  &  \\
\hline

\multirow{9}{*}{$D^{*}NN$}
%---------------- J^P = 0^- ----------------
& \multirow{3}{*}{$0^-$}
& 1.35 & --   & --   & 0.03 & 5.97 & --   & --   & -- & -- & 31.60 & 29.37 & 1.19 & 1.11 \\
&      & 1.40 & --   & --   & 0.57 & 4.47 & --   & --   & -- & -- & 42.71 & 40.48 & 1.09 & 1.00 \\
&      & 1.45 & --   & --   & 1.40 & 3.32 & --   & --   & -- & -- & 53.35 & 51.12 & 1.03 & 0.93 \\
\cline{2-15}

%---------------- J^P = 1^- (two branches) ----------------
& \multirow{3}{*}{$1^-$}
& 1.35 & -- & -- & 0.03 & 5.97 & 7.33 & 1.80 & -- & --
& \begin{tabular}[c]{@{}c@{}} 58.48\\ 18.44 \end{tabular}
& \begin{tabular}[c]{@{}c@{}} 51.15\\ 11.11 \end{tabular}
& \begin{tabular}[c]{@{}c@{}} 1.14\\ 1.46 \end{tabular}
& \begin{tabular}[c]{@{}c@{}} 1.04\\ 1.44 \end{tabular} \\
&      & 1.40 & -- & -- & 0.57 & 4.47 & 9.75 & 1.62 & -- & --
& \begin{tabular}[c]{@{}c@{}} 66.84\\ 29.69 \end{tabular}
& \begin{tabular}[c]{@{}c@{}} 57.09\\ 19.94 \end{tabular}
& \begin{tabular}[c]{@{}c@{}} 1.12\\ 1.28 \end{tabular}
& \begin{tabular}[c]{@{}c@{}} 1.01\\ 1.20 \end{tabular} \\
&      & 1.45 & -- & -- & 1.40 & 3.32 & 12.33 & 1.48 & -- & --
& \begin{tabular}[c]{@{}c@{}} 75.72\\ 45.22 \end{tabular}
& \begin{tabular}[c]{@{}c@{}} 63.39\\ 32.89 \end{tabular}
& \begin{tabular}[c]{@{}c@{}} 1.08\\ 1.14 \end{tabular}
& \begin{tabular}[c]{@{}c@{}} 0.98\\ 1.04 \end{tabular} \\
\cline{2-15}

%---------------- J^P = 2^- ----------------
& \multirow{3}{*}{$2^-$}
& 1.35 & -- & -- & -- & -- & 7.33 & 1.80 & -- & -- & 29.14 & 21.81 & 1.27 & 1.19 \\
&      & 1.40 & -- & -- & -- & -- & 9.75 & 1.62 & -- & -- & 37.91 & 28.16 & 1.16 & 1.06 \\
&      & 1.45 & -- & -- & -- & -- & 12.33 & 1.48 & -- & -- & 49.35 & 37.02 & 1.07 & 0.96 \\

\hline\hline
\end{tabular}
\end{table*}

%\bibliography{ref}
%

\end{document}